\documentstyle[12pt]{article}
\title{
TIME EVOLUTION IN THE EXTERNAL FIELD: THE UNITARITY PARADOX
}

\author{\large  O.Yu.Shvedov \\
{\it
Sub-Dept. of Quantum Statistics and Field Theory},\\
{\it Dept. of Physics, Moscow State University},\\
{\it 119899, Moscow, Vorobievy Gory, Russia}
}
\date{}
\begin{document}

\def\qp{
\mathrel{\mathop{\bf x}\limits^2},
\mathrel{\mathop{-i\frac{\partial}{\partial {\bf x}}}\limits^1} 
}

\maketitle

\begin{flushright}
hep-th/0002108
\end{flushright}

\begin{abstract}

One of the axioms of quantum field theory is the property of
unitarity of the evolution operator. However, if one considers the
quantum electrodynamics in the external field in the leading order
of perturbation   theory,   one   will   find   that   the   evolution
transformation is a non-unitary canonical transformation  of  creation
and annihilation operators.  This observation was one of the arguments
for the hypothesis that one should choose different representations of
the canonical  commutation  relations  at different moments of time in
the exact quantum field theory.  In this paper  the  contradiction  is
analyzed for  the  case  of  a simple quantum mechanical model being an
analog of the leading order of the large-N field theory.  On  the  one
hand,  this  model  is  renormalized with the help of the constructive
field theory methods; the Hilbert space and unitary evolution operator
are constructed. On the other hand, the leading order of the evolution
transformation  in  the  strong  external  field  is   shown   to   be
non-unitary.  Thus,  unitarity of evolution in the exact theory is not
in contradiction with non-unitarity of the approximate theory.

\end{abstract}

\footnotetext{e-mail:  shvedov@qs.phys.msu.su}

\newcounter{eqn}
\def\lab{\refstepcounter{eqn}\eqno(\arabic{eqn})}
\def\l#1{\lab\label{#1}}
\def\r#1{(\ref{#1})}
\def\c#1{\cite{#1}}
\def\i#1{\bibitem{#1}}

\newpage

\section{Introduction}

One of the main postulates of the axiomatic
relativistic quantum field theory in the Wightman  approach  \c{W1,SW}
(for a review see \c{1}) is
the following.  The  Poincare  transformations   should   be   unitary
operators in  the Hilbert state space.  For example,  this postulate
should be correct for the operator of  time  evolution.  However,  the
check of  this  axiom for realistic models of QFT is difficult,  since
the models are usually constructed in the perturbative approach only.

On the other hand,  one can investigate the (3+1)-dimensional spinor
QED in  the strong external classical electromagnetic field \c{2,3,4,J}.
It happens that even in the leading order of perturbation  theory  the
creation and  annihilation  operators at different moments of time are
related with the help of {\it non-unitary} Bogoliubov canonical
transformation. This  means  that  for constructing QFT
in the  external   field,   it   is   necessary   to   use   different
representations of   the  canonical  commutation  relations  (CCR)  at
different moments of time.

This observation  implies  that   one   can   expect   that   in   the
non-perturbative QFT    one    should    also    consider    different
representations of CCR at different moments of time \c{6},  while  the
time translation  (evolution)  is {\it not} an unitary operator in the
Hilbert space but transformation connecting different  representations
of CCR.  This  suggestion  is  in  contradiction  with  the  Wightman
axiomatic approach.  However, it is in agreement with the more general
algebraic approach  developed  by  Haag  and Kastler \c{7} (for recent
reviews of the algebraic approach see \c{Buchholz, Schroer}).

It is shown in this paper that non-unitarity of the evolution operator
in the external field in the leading order of perturbation theory does
{\it not} contradict to the unitarity axiom of the "exact" theory. The
simple exactly solvable model is considered in this paper. The Hilbert
state  space and {\it unitary} evolution operator are constructed with
the help of the Bogoliubov $S$-matrix approach \c{8,9}. This model can
be  also  considered in the strong external field in the leading order
of perturbation theory:  the corresponding states are constructed. The
evolution    transformation   is   a   {\it   non-unitary}   canonical
transformation.

This model corresponds to the (0+1)-dimensional
"field" $Q(t)$ interacting with infinite
number of "fields" $Q_k(t)$. The action of the model is
$$
S = \int dt
 [L_Q + \sum_{k=1}^{\infty}
\left(\frac{\dot{q}_k^2}{2} - \frac{\Omega_k^2q_k^2}{2}\right)
- g\sum_{k=1}^{\infty} \mu_k q_k Q],
\l{1}
$$
where
$$
L_Q =    \sum_{s=0}^l    (-1)^{s+1}    z_s     (Q^{(s)})^2,     \qquad
Q^{(s)}=\frac{d^sQ}{dt^s},
$$
while $\mu_k$  are  some  coefficients.  As  $k\to\infty$,  the set of
numbers $\Omega_k$ tends to infinity.
If $\sum_{k=1}^{\infty}  \frac{\mu_k^2}{\Omega_k^2}  =  \infty$,   the
problem of divergences arises. However, if
$\sum_{k=1}^{\infty} \frac{\mu_k^2}{\Omega_k^m}  =  \infty$  for  some
$m$, the divergences can be renormalized.

Note that the model  \r{1}  is  a  quantum-mechanical  analog  of  the
large-$N$ theory $\Phi\varphi^a\varphi^a$.
Large-$N$ conception   is   very   useful  in  QFT:  one  can  perform
resummation of  Feynman  graphs,  making  use  of  the   diagram   and
functional-integral techniques \c{W,CJT,CJP}.  The problem of external
field (the back reaction)
can be also investigated in the large-$N$ theory \c{CM,CM1}.

Consider the theory of $N$ fields $\varphi^a$
interacting with   the   field   $\Phi$   in  the  $(d+1)$-dimensional
space-time. The Lagrangian of the theory is
$$
\begin{array}{c}
{\cal L} = \sum_{a=1}^N :\left(
\frac{1}{2}\partial_{\mu} \varphi^a\partial_{\mu} \varphi^a
- \frac{\mu^2}{2} \varphi^a \varphi^a \right):
+ \frac{z}{2}\partial_{\mu} \Phi\partial_{\mu} \Phi
- \frac{M^2}{2} \Phi^2 \\
- \frac{g}{\sqrt{N}} :\left(
\sum_{a=1}^N \varphi^a \varphi^a \right):\Phi
\end{array}
$$
Analogously to  \c{B2}  (see also \c{JP,JL,KR,Y}),
introduce the "collective fields"
being the operators  of  creation   and
annihilation of pairs of particles
$$
A^{\pm}_{{\bf k}{\bf p}} = \frac{1}{\sqrt{2N}} \sum_{a=1}^N
b^{\pm a}_{\bf k} b^{\pm a}_{\bf p},
$$
where $b^{\pm a}_{\bf k}$ is a creation-annihilation operator
of the particle with momentum
${\bf k}$, which corresponds to the field $\varphi^a$.

We will consider the states of the $N$-field theory  which  depend  on
the large parameter $N$ as follows
$$
\sum_n \int
d{\bf k}_1d{\bf p}_1 ...
d{\bf k}_nd{\bf k}_n
A^+_{{\bf k}_1{\bf p}_1}... A^+_{{\bf k}_n{\bf p}_n}
\chi^n_{{\bf k}_1{\bf p}_1...{\bf k}_n{\bf p}_n} \Psi,
\l{z3}
$$
with regular as $N\to\infty$ coefficient functions
$\chi^n$
and such vector $\Psi$ that does not contain
the particles corresponding to the fields
$\varphi^a$.

Note that operators of the form
$$
\int d{\bf k} d{\bf p} \frac{1}{\sqrt{N}}
\sum_{a=1}^N b_{{\bf k}}^{+a} b_{\bf p}^{-a} \varphi_{{\bf k}{\bf p}}
$$
multiply the norm of the state \r{z3} by the quantity
$O(N^{-1/2})$. Therefore, they can be neglected
as $N\to\infty$. In this approximation
$$
[A^-_{{\bf k}_1{\bf p}_1};
A^+_{{\bf k}_2{\bf p}_2}]\simeq
\frac{1}{2}
(
\delta_{{\bf k}_1{\bf k}_2}
\delta_{{\bf p}_1{\bf p}_2} +
\delta_{{\bf k}_1{\bf p}_2}
\delta_{{\bf k}_2{\bf p}_1}
).
$$

Consider the free Hamiltonian
$H_0  =  \int
d{\bf k} \omega_{\bf k}
\sum_{a=1}^N b_{\bf k}^{+a}b_{\bf k}^{-a}$,
where $\omega_{\bf k}= \sqrt{{\bf k}^2+\mu^2}$.
If we consider the states of the form \r{z3} only, it coincides with
the operator
$$
\int d{\bf   k}   d{\bf   p}  A^+_{{\bf  k}{\bf  p}}  (\omega_{\bf  k}
+ \omega_{\bf p})  A^-_{{\bf  k}{\bf  p}},
$$
The operator
$\frac{1}{\sqrt{N}}\sum_{a=1}^N  \varphi^a({\bf x})
\varphi^a({\bf x})$ is approximately equal to
$$
\frac{\sqrt{2}}{(2\pi)^d}
\int \frac{d{\bf k}}{\sqrt{2\omega_{\bf k}}}
\frac{d{\bf p}}{\sqrt{2\omega_{\bf p}}}
(A^+_{{\bf  k}{\bf  p}}e^{-i({\bf k}+{\bf p}){\bf x}}
+ A^-_{{\bf  k}{\bf  p}}e^{i({\bf k}+{\bf p}){\bf x}} ).
$$
The leading  order  for  the  Hamiltonian  in  $1/N$  is  analogous to
eq.\r{1}:
$$
\begin{array}{c}
H= \int d{\bf k}d{\bf p}
 A^+_{{\bf  k}{\bf  p}}  (\omega_{\bf  k}
+ \omega_{\bf p})  A^-_{{\bf  k}{\bf  p}}
+ \int d{\bf x}
\left(
\frac{1}{2z}\Pi^2({\bf x}) +
\frac{z}{2} (\nabla \Phi)^2({\bf x})
+ \frac{M^2}{2} \Phi^2
\right)
\\
+ \frac{\sqrt{2}g}{(2\pi)^d}
\int d{\bf x}
\left[
\int \frac{d{\bf k}}{\sqrt{2\omega_{\bf k}}}
\frac{d{\bf p}}{\sqrt{2\omega_{\bf p}}}
(A^+_{{\bf  k}{\bf  p}}e^{-i({\bf k}+{\bf p}){\bf x}}
+ A^-_{{\bf  k}{\bf  p}}e^{i({\bf k}+{\bf p}){\bf x}})
\right] \Phi({\bf x}).
\end{array}
\l{zz4}
$$
The index
$k$ is substituted by $({\bf k},{\bf p})$, the sums are substituted by
integrals.
Eq.\r{zz4} can be also
obtained from the third-quantized approach
\c{10,11}.

Investigation of  the models of the type \r{1} allows us to understand
the difficulties of the quantum field theory in the external field.

This paper  is  organized  as  follows.  In section 2 we construct the
Hilbert state space and unitary evolution operator for the model \r{1}
which occurs  to be renormalizable.  Section 3 deals with constructing
special states of the model  \r{1}  which  correspond  to  the  strong
classical external  field.  Time  evolution  in  the  obtained quantum
theory in the external field is shown to be  a  non-unitary  canonical
transformation. Section 4 is devoted to the analysis of the paradox.

\section{Construction of   the   model:   evolution   as   an  unitary
transformation}

In this section the quantized model \r{1}
is constructed. We  show  that  the
evolution operator is a well-defined unitary transformation
in the Hilbert state space.

\subsection{The Bogoliubov $S$-matrix and unitarity of evolution}

\subsubsection{Conditions on the Hamiltonian and on the state}

Models of  the  constructive  field  theory  are usually formulated as
follows \c{13,14}.  Instead of the Schrodinger equation  obtained  by
the formal quantization procedure,
$$
i \dot{\Psi}_t = [\hat{H}_0+g\hat{H}_1] \Psi_t,
\l{14}
$$
where $\hat{H}_0$ is a free Hamiltonian, $\hat{H}_1$ is an interaction
containing UV divergences, one considers the regularized equation.
The Hamiltonian
$\hat{H}_1$ is substituted by the regular operator
$\hat{H}_1^{\Lambda}$ depending on the cutoff parameter $\Lambda$.  At
finite values   of   $\Lambda$   this   operator   does   not  contain
UV-divergences. As $\Lambda\to\infty$, the regularized expression for
the Hamiltonian  formally  tends  to $\hat{H}_1$.  Usually, it
is necessary to add the counterterms $\hat{H}^{\Lambda}_{ct}(g)$
to the Hamiltonian. The evolution equation reads,
$$
i\dot{\Psi}^{\Lambda}_t = \hat{H}^{\Lambda}\Psi^{\Lambda}_t =
[\hat{H}_0 + g\hat{H}^{\Lambda}_1+
\hat{H}_{ct}^{\Lambda}(g)]\Psi_t^{\Lambda}.
\l{15}
$$
In the  $S$-matrix  approach  one  imposes  the  conditions   on   the
dependence of  the  counterterms  on  the cutoff parameter in order to
obtain the finite $S$-matrix. Within the perturbation framework, it is
possible: the     well-known     Bogoliubov-Parasiuk     theorem    on
renormalizability of QFT if proved. Contrary to the $S$-matrix approach,
in  order  to  eliminate divergences from the equations of motion,
it is {\it  not}  sufficient  to impose conditions   on   the
counterterms. Even   in  the  tree  Feynman  graphs  the  Stueckelberg
divergences arise \c{15} when one investigates the processes at finite
time intervals  (like  emission  of  the  virtual  photon  by a single
electron \c{15}).  To  eliminate  the  Stueckelberg  divergences,  one
should also  impose  the  conditions on the dependence on $\Lambda$ of
the initial condition for eq.\r{15}. The problem of elimination of the
Stueckelberg divergences   for  the  leading  order  of  semiclassical
expansion was investigated in \c{MS1,Baacke,S2}.

In the  constructive   field   theory \c{13,16}
one   usually   chooses   such
$t$-independent unitary  transformation  $T_{\Lambda}$  (singular   as
$\Lambda\to\infty$) that the following requirement is satisfied.
Suppose that the initial condition for eq.\r{15} depends on  $\Lambda$
as
$\Psi^{\Lambda}_0 = T_{\Lambda} \Phi_0^{\Lambda}$,
where the vector
$\Phi^{\Lambda}_0$ has a strong limit as
$\Lambda\to \infty$.  Then  the  solution  to  the  Cauchy problem for
eq.\r{15} should have an analogous form:
$$
\Psi^{\Lambda}_t = T_{\Lambda} \Phi^{\Lambda}_t.
\l{16}
$$
with regular as $\Lambda\to\infty$ vector $\Phi^{\Lambda}_t$,
$\Phi^{\Lambda}_t \to_{\Lambda\to\infty} \Phi_t$. The vector
$\Phi_t$ can be viewed as a ``renormalized'' state.
The operator transforming the state $\Phi_0$
to the state $\Phi_t$ is regular as
$\Lambda\to\infty$,
$$
U_t = s-\lim_{\Lambda\to\infty} U^t_{\Lambda}
= \lim_{\Lambda\to\infty} (T_{\Lambda})^{-1}
\exp[-i\hat{H}_{\Lambda}t]T_{\Lambda}
\l{17}
$$
It can  be  viewed  as  a renormalized evolution operator in the model
\r{14}.

Note that the evolution operator $\exp[-i\hat{H}_{\Lambda}t]$  may  be
singular as  $\Lambda\to\infty$ while the renormalized operator \r{17}
may be regular.

Consider the arbitrary observable $O$ corresponding  to  the  operator
$O_{\Lambda}$ in the regularized theory.  In the representation \r{16}
it can be written as:
$$
T_{\Lambda}^{-1} \hat{O}_{\Lambda} T_{\Lambda}.
\l{b4}
$$
If eq.\r{b4} possesses the limit $\Lambda\to\infty$, one can talk about
the {\it time-independent} representation of the observable $O$ in the
''renormalized'' state   space.  In  particular,  this  is  a  way  to
construct a time-independent non-Fock representation of the  canonical
commutation relation.

\subsubsection{The Bogoliubov approach}

To  construct the operator
$T_{\Lambda}$, let us use the axiomatic Bogoliubov
approach \c{8}  (see  also \c{9}) based on the conception of switching
on the interaction. In this approach one considers the analog of the
model \r{14}  with the {\it time-dependent} coefficient of interaction
$g_t=g(t)$ instead of the  case  of  the  constant  interaction.  This
generalization of the model seems to be a complication.  However, if
one considers the case of the smooth function $g$  which  is  non-zero
only on  the  finite time interval,  the $S$-matrix will be regular as
$\Lambda \to\infty$,  contrary to the evolution operator which can  be
viewed as  $S$-matrix  corresponding  to the discontinuous function $g$
being constant at $t\in(t_1,t_2)$ and  zero  at  $t\notin  (t_1,t_2)$.
The $S$-matrix  viewed  as  a  functional  on  the smooth function $g$
vanishing at sufficiently large  $|t|$  is  the  main  notion  of  the
Bogoliubov axiomatic approach.
One  can  determine  the renormalized evolution
operator in terms of $S$-matrix \c{9}.

After regularization and  renormalization  the  Bogoliubov  $S$-matrix
takes the form
$$
S_{\Lambda}[g]=Texp(-i\int_{-\infty}^{\infty} e^{i\hat{H}_0\tau}
(g(\tau)\hat{H}_1^{\Lambda}+\hat{H}_{ct}^{\Lambda}[\tau,g(\cdot)])
e^{-i\hat{H}_0\tau}d\tau).
\l{19}
$$
It transforms the initial condition for the equation
$$
i \frac{d\tilde{\Phi}^t_{\Lambda}}{dt} =
 e^{i\hat{H}_0 t}
(g(t)\hat{H}_1^{\Lambda}+\hat{H}_{ct}^{\Lambda}[t,g(\cdot)])
e^{-i\hat{H}_0t}\tilde{\Phi}^t_{\Lambda}
\l{20}
$$
as $t=-\infty$ to the solution of this equation
at  $t=+\infty$,
$S_{\Lambda}(g)\tilde{\Phi}^{-\infty}=\tilde{\Phi}^{+\infty}$.
Note the function $g(t)$ is zero at $|t|>T$.

The counterterms
$\hat{H}_{ct}^{\Lambda}[t,g(\cdot)])$
depending on the function
$g$ and its derivatives at time moment
$t$ are   chosen  in  order  to  make  the  $S$-matrix  regular.  More
precisely, for smooth functions $g(t)$ the
$S$-matrix should have a strong limit as $\Lambda\to\infty$.

In the  interaction  representation for finite values of $\Lambda$ the
evolution operator coincides with the Bogoliubov $S$-matrix \r{19} if
$g(\tau)=g$  at
$\tau \in [t_1,t_2]$ and $g(\tau)=0$ at $\tau\notin [t_1,t_2]$.
Namely, the substitution
$\Psi_{\Lambda}^t=e^{-i\hat{H}_0t}
\tilde{\Phi}_{\Lambda}^t$
transforms eq.\r{15}
to the form \r{19}, so that
$$
e^{-i\hat{H}_{\Lambda} (t_2-t_1)} = e^{-i\hat{H}_0t_2}
S_{\Lambda} [gI_{t_1t_2}(\cdot)] e^{i\hat{H}_0t_1},
$$
where
$I_{t_1t_2}(t) = 1$ as $t\in (t_1,t_2)$
¨
$I_{t_1t_2}(t) = 0$ as  $t\notin (t_1,t_2)$.
However, because of the
Stueckelberg divergences  \c{15,9} the strong limit
of the $S$-matrix as $\Lambda\to\infty$ for the case of
a discontinuous function $g$, in general, does not exist.

Consider the function $\xi_-(\tau)$ which switches on
from 0 to 1 at $-T_1<\tau<-T_2$, $-T_2<0$, is equal to 1
at $-T_2<\tau<0$ and 0 at  $\tau>0$ and $\tau<-T_1$ (see fig.1).

Choose the unitary operator
$T_{\Lambda}$ to be the following:
$$
T_{\Lambda} = S_{\Lambda} [g\xi_-(\cdot)].
\l{20*}
$$
Consider also the function
$\xi_+(\tau)=\xi_-(-\tau)$ and operator
$v_t$ of shifting the argument $\tau$:
$v_tg(\tau)=g(\tau+t)$.
The operator $U^t_{\Lambda}$ entering to eq.\r{17} takes the form:
$$
U_{\Lambda}^{t_2-t_1} = S^+_{\Lambda} [g\xi_-(\cdot)]
e^{-i\hat{H}_0t_2}
S_{\Lambda}[gI_{t_1t_2}(\cdot)] e^{i\hat{H}_0t_1}
S_{\Lambda}[g\xi_-(\cdot)].
$$
The property
$S_{\Lambda}[v_tg] = e^{i\hat{H}_0t} S_{\Lambda}[g] e^{-i\hat{H}_0t}$
and unitarity of the operator
$S_{\Lambda}[g\xi_+(\cdot)]$ imply that
$$
\matrix{
U_{\Lambda}^{t_2-t_1} = e^{-i\hat{H}_0t_2}
S_{\Lambda}^+[gv_{t_2}\xi_-(\cdot)]
S_{\Lambda}^+[gv_{t_2}\xi_+(\cdot)]
\cr \times
S_{\Lambda}[gv_{t_2}\xi_+(\cdot)]
S_{\Lambda}[gI_{t_1t_2}(\cdot)]
S_{\Lambda}[gv_{t_1}\xi_-(\cdot)]
e^{i\hat{H}_0t_1}.
}
\l{21}
$$
Denote as
$\xi_{t_1t_2}$ the smooth function of the form
$$
\xi_{t_1t_2}= v_{t_1}\xi_- + I_{t_1t_2} + v_{t_2}\xi_+,
\qquad t_1\le t_2.
$$
The operator \r{21} can be presenter as
$$
U_{\Lambda}^{t_2-t_1} = e^{-i\hat{H}_0t_2}
S^+_{\Lambda}[g\xi_{t_2t_2}(\cdot)]
S_{\Lambda}[g\xi_{t_1t_2}(\cdot)]
e^{i\hat{H}_0t_1}.
$$
Thus, the  operator  \r{21}  is  expressed  via  the  values  of   the
Bogoliubov $S$-matrix  functional  on  the  smooth functions $g$ which
vanish at $|t|>T$ for some $T$.

In the next subsection we show that the operators
$S_{\Lambda}[g(\cdot)]$ and $S^+_{\Lambda}[g(\cdot)]$ have
strong limits as $\Lambda\to\infty$.

This will imply that the {\it renormalized} evolution operators \r{17}
exist.

Note that two operators
$T_{\Lambda}$ corresponding to different functions
of switching the interaction
$\xi^{(1)}_-$ and  $\xi^{(2)}_-$ lead to equivalent representations of
the observables since the unitary operator
$
S^+_{\Lambda}[g\xi_-^{(1)}]
S_{\Lambda}[g\xi_-^{(2)}]
=
S^+_{\Lambda}[g(\xi_-^{(1)}+\xi_+^{(1)})]
S_{\Lambda}[g(\xi_-^{(2)}+\xi_+^{(1)})]
$
has a (strong) limit as $\Lambda\to\infty$.

\subsection{Construction of the Bogoliubov $S$-matrix}

\subsubsection{Regularization and canonical quantization}

Consider the  canonical  quantization  of  the model \r{1}.  Since the
Lagrangian contains higher derivatives,  the classical Hamiltonian  of
the model depends \c{17} on the coordinates
$V_0=Q$,
$V_1=\dot{Q}$, ...,
$V_{l-1}=Q^{(l-1)}$ and canonically conjugated momenta
$P_0$, $P_1$,
..., $P_{l-1}$, as well as on the coordinates and momenta
$q_k$ and $p_k$. The classical Hamiltonian function has the form:
$$
H = H_q + H_Q + g \sum_{k=1}^{\infty} \mu_k q_k Q,
\l{22}
$$
where
$$
\matrix{
H_q = \sum_{k=1}^{\infty} \left(
\frac{p_k^2}{2} + \frac{\Omega_k^2q_k^2}{2}
\right),
\cr
H_Q = \frac{(-1)^{l+1}P_{l-1}^2}{2z_l}
+ \sum_{s=0}^{l-2} P_sV_{s+1} +
\sum_{s=0}^{l-1} \frac{(-1)^s}{2} z_s V_s^2.
}
$$
Under the  canonical  quantization  procedure,  the  coordinates   and
momenta are associated with the operators
$\hat{V}_0,...,\hat{V}_{l-1}$,
$\hat{P}_0,...,\hat{P}_{l-1}$,
$\hat{q}_k$, $\hat{p}_k$ obeying the canonical  commutation  relations
(CCR):
$$
[\hat{V}_m, \hat{P}_k] = i\delta_{mk},
\qquad
[\hat{q}_k, \hat{p}_l] = i\delta_{kl}.
\l{o1}
$$
Other commutators vanish.

In order to avoid the divergences at the intermediate  stages  of  the
analysis of  the model,  introduce the regularization.  The quantities
$\mu_k$ are substituted by
$$
\mu_k^{\Lambda} = \mu_k; \quad k<\Lambda,
\qquad \mu_k^{\Lambda} = 0, \quad k\ge\Lambda.
$$
where $\Lambda$ is a cutoff parameter.
We will show that the  Bogoliubov
$S$-matrix is regular as $\Lambda\to \infty$ if
the counterterm
$$
H_{ct}^{\Lambda} =
\sum_{s=0}^{l-1} \frac{(-1)^s}{2}\delta z_s V_s^2.
$$
is added to the Hamiltonian \r{22}.

In the  cutoffed  theory  one  can  use the Fock representation of CCR
\r{o1}. The  operators  $\hat{p}_k$  and $\hat{q}_k$ are presented via
the creation and annihilation operators $\hat{a}^{\pm}_k$
in the Fock space:
$$
\hat{q}_k = \frac{\hat{a}^+_k + \hat{a}^-_k }{\sqrt{2\Omega_k}};
\hat{p}_k = i \sqrt{\frac{\Omega_k}{2}}(\hat{a}^+_k - \hat{a}^-_k),
\l{a2*}
$$
They obey the following relations:
$$
[\hat{a}_k^-,\hat{a}_m^+]
= \delta_{km}, \quad [\hat{a}^{\pm}_k, \hat{a}^{\pm}_l] =0.
$$
Remind that the Fock space $\cal F$ is a space of sets
$$
\Psi=(\Psi_0,\Psi_1(k_1),\Psi_2(k_1,k_2),...)
$$
of symmetric with respect to $k_1,...,k_n$ functions
$\Psi_n(k_1,...,k_n)$,  $k_1,...,k_n=1,2,3,...$,
such that the series
$$
\sum_{n=0}^{\infty}                        \sum_{k_1...k_n=1}^{\infty}
|\Psi_n(k_1,...,k_n)|^2 <\infty
\l{a3a}
$$
converges. The operators $\hat{a}_k^{\pm}$ act in the Fock space as
$$
(\hat{a}_k^+\Psi)_n(k_1,...,k_n) =      \sum_{j=1}^n      \frac{1}{\sqrt{n}}
\delta_{kk_j} \Psi_{n-1}(k_1,...,k_{j-1},k_{j+1},...,k_n),
$$
$$
(\hat{a}_k^-\Psi)_{n-1}(k_1,...,k_{n-1}) = \sqrt{n} \Psi_n(k,k_1,...,k_{n-1}).
$$

The operators   $\hat{V}_i$   act   in   the   space   of    functions
$\psi(V_0,...,V_{l-1})$ from   $L^2({\bf   R}^l)$   as   operators  of
multiplication by $V_i$,  while  $\hat{P}_i=-i\frac{\partial}{\partial
V_i}$.

Choose, as  usual,  the  space  ${\cal  F}\otimes L^2({\bf R}^l)$ as a
state space  of  the  composed  system;  the  operators
$\hat{P}_i$, $\hat{V}_i$ and $\hat{a}_k^{\pm}$
are extended as
$\hat{P}_i \otimes    1$,    $\hat{V}_i\otimes    1$   and   $1\otimes
\hat{a}_k^{\pm}$ correspondingly.

Consider eq.\r{20}  for  this   model.   Instead   of   the   operator
$e^{iH_0t}(\hat{H}-\hat{H}_0)e^{-iH_0t}$ entering  to  the  right-hand
side of eq.\r{20},  where $\hat{H}_0= \hat{H}_Q+ \hat{H}_q$,  consider
the operator of the form
$$
e^{i\hat{H}_qt} (\hat{H}-\hat{H}_q) e^{-i\hat{H}_qt} =
\hat{H}_Q + \hat{H}_{ct}^{\Lambda} + g \sum_{k=1}^{\infty}
\mu_k^{\Lambda} \hat{Q}
\frac{\hat{a}_k^+e^{i\Omega_kt} + \hat{a}_k^-e^{-i\Omega_kt}}
{\sqrt{2\Omega_k}}.
\l{23}
$$
Denote by  $U^{\Lambda}_{t_1t_2}$  the  evolution  operator  for   the
Hamiltonian \r{23}  transforming  the  initial  condition  at  for the
equation
$$
i\frac{d}{dt} \Phi_{\Lambda}^t =
e^{i\hat{H}_qt} (\hat{H}-\hat{H}_q) e^{-i\hat{H}_qt}\Phi_{\Lambda}^t
$$
at $t=t_1$ to the solution of this equation at $T=t_2$,
$\Phi_{\Lambda}^{t_2} = U^{\Lambda}_{t_1t_2}\Phi_{\Lambda}^{t_1}$. The
evolution operator  $\tilde{U}^{\Lambda}_{t_1t_2}$  for  eq.\r{20}  is
related with $U^{\Lambda}_{t_1t_2}$ as
$$
\tilde{U}^{\Lambda}_{t_1t_2}=
e^{iH_Qt_2}
U^{\Lambda}_{t_1t_2} e^{-iH_Qt_1}.
$$
Let $g=g(t)=g_t$ be a function vanishing at $|t|>T$.
Then the   Bogoliubov   $S$-matrix   coincides   with   the   operator
$\tilde{U}^{\Lambda}_{-TT}$ and  can  be  expressed   then   via   the
evolution operator
${U}^{\Lambda}_{-TT}$
for the Hamiltonian \r{23}.  We will show that this
operator has a strong limit as  $\Lambda\to\infty$.  This  will  imply
that the Bogoliubov $S$-matrix and the renormalized evolution operator
for the $g=const$-case are well-defined as $\Lambda=\infty$.

\subsubsection{
Renormalization of divergences in classical equations}

It is well-known \c{5,9}
that quantum theories with quadratic  Hamiltonians  are  specified  by
their classical  analogs.  Consider  the  divergences in the classical
version of the model \r{23}.  Equations  of  motion  for  the  quantum
Heisenberg operators  coincide  with  the classical equations and have
the form:
$$
i\dot{a}_k^- = g \frac{\mu_k}{\sqrt{2\Omega_k}} Q e^{i\Omega_k t},
\qquad
-i\dot{a}_k^+ = g \frac{\mu_k}{\sqrt{2\Omega_k}} Q e^{-i\Omega_k t};
\l{24}
$$
$$
\matrix{
\quad \dot{V}_j = V_{j+1}, j=\overline{0,l-1};
\dot{V}_{l-1} = (-1)^{l-1} \frac{P_{l-1}}{z_l};
\cr
-\dot{P}_j = (-1)^j (z_j+\delta z_j)V_j + P_{j-1}, j=\overline{1,l-1};
\cr
-\dot{P}_0 = (z_0+\delta z_0)Q +
g\sum_{k=1}^{\infty}
\mu_k
\frac{a_k^+e^{i\Omega_kt}+ a_k^-e^{-i\Omega_kt}}
{\sqrt{2\Omega_k}}.
}
\l{25}
$$
Eqs.\r{25} can be presented as
$$
\sum_{s=0}^l (z_s Q^{(s)})^{(s)} + \sum_{s=0}^{l-1}
(\delta z_s Q^{(s)})^{(s)} +
g\sum_{k=1}^{\infty}
\mu_k
\frac{a_k^+e^{i\Omega_kt}+ a_k^-e^{-i\Omega_kt}}
{\sqrt{2\Omega_k}} = 0,
\l{26}
$$
while integration of eqs.\r{24} gives us the following relations:
$$
a_k^{\pm}(t) = a_k^{\pm}(-\infty)
\pm i \frac{\mu_k}{\sqrt{2\Omega_k}}
\int_{-\infty}^t d\tau g_{\tau} Q_{\tau} e^{\mp i \Omega_k \tau}.
\l{27}
$$
After substitution of formula \r{27} and integration by parts
$2l$ times eq. \r{26} is transformed to the following form:
$$
\matrix{
\sum_{s=0}^l (z_s Q_t^{(s)})^{(s)} +
\sum_{k=1}^{\infty}
g_t \mu_k
\frac{a_k^+(-\infty)e^{i\Omega_kt}+ a_k^-(-\infty)
e^{-i\Omega_kt}} {\sqrt{2\Omega_k}}
\cr
+ (-1)^{l+1} g_t \sum_{k=1}^{\infty}
\frac{\mu_k^2}{\Omega_k}
\int_{-\infty}^t d\tau (g_{\tau} Q_{\tau})^{(2l)}
\frac{sin(\Omega_k(t-\tau))}{\Omega_k^{2l}} =
-(\hat{B}_1+\hat{B}_2)Q(t)
}
\l{28}
$$
where
$$
\hat{B}_1 = \sum_{s=0}^{l-1} \frac{\mu_k^2}{\Omega_k^{2s+2}}
(-1)^{s+1} g_t \frac{d^{2s}}{dt^{2s}} g_t;
\qquad \hat{B}_2=\sum_{s=0}^{l-1} \frac{d^s}{dt^s} \delta z_s
\frac{d^s}{dt^s}.
$$
Check that under some choice of the counterterms
$\hat{B}_2+\hat{B}_1=0$. Note that the operator
 $\hat{B}_1$  is Hermitian and polynomial in
$d/dt$. The degree of the polynomial is
$2l-2$, so that
$\hat{B}_1= \sum_{j=0}^{2l-2} b_j(t) \frac{d^j}{dt^j}$.
Choose
$$
\hat{B}^{(1)}_2 = \frac{d^{l-1}}{dt^{l-1}} b_{2l-2}(t)
\frac{d^{l-1}}{dt^{l-1}} = b_{2l-2}
\frac{d^{2l-2}}{dt^{2l-2}} + (l-1) \dot{b}_{2l-2}
\frac{d^{2l-3}}{dt^{2l-3}}+...
$$
One has
$$
\hat{B}^{+}_1 =
b_{2l-2}
\frac{d^{2l-2}}{dt^{2l-2}} +
((2l-2) \dot{b}_{2l-2} - b_{2l-3})
\frac{d^{2l-3}}{dt^{2l-3}}+...
$$
and  $b_{2l-3}=(l-1)\dot{b}_{2l-2}$,  so that the operator
$\hat{B}_1-\hat{B}_2^{(1)}$
contains the derivative
$d/dt$ in degrees no higher than
$2l-4$.
Applying to the operator
$\hat{B}_1-\hat{B}_2^{(1)}$
an analogous procedure
$l-2$ times, one constructs such operator
$\hat{B}_2=\hat{B}_2^{(1)}+...+\hat{B}_2^{(l-2)}$, that
$\hat{B}_2=-\hat{B}_1$.

Eq.\r{28} does not contain singularities if
$$
\sum_k \frac{\mu_k^2}{\Omega_k^{2l+1}} < \infty.
\l{13}
$$
Existence, uniqueness and smoothness of the  solution  of  the  Cauchy
problem for  this  equation  are  corollaries of the general theory of
integral equations (see,  for example,  \c{18}).  Thus,  the classical
theory does not contain singularities as
$\Lambda\to\infty$, provided  that  the  considered  counterterms  are
added and condition \r{13} is satisfied.

\subsubsection{Regularity of the
Bogoliubov $S$-matrix}

Let us check that under condition \r{13} the evolution operator
$U^{\Lambda}_{-TT}$ for  the  time  interval  $(-T,T)$  is  regular as
$\Lambda\to\infty$ in the theory with Hamiltonian \r{23}.
For simplicity of the notations, the index $\Lambda$ will be omitted.
Represent the operators
$\hat{Q}=V_0$,
$\hat{V}_s$,  $\hat{P}_s$ via creation and annihilation operators
$
\hat{B}_k^{\pm} = \frac{\hat{V}_k\mp i \hat{P}_k}{\sqrt{2}}$;
denote $\hat{B}^{\pm}_{l+k}=\hat{a}_k^{\pm}$.
The Hamiltonian \r{23} is quadratic in creation-annihilation operators:
$$
H= \sum_{ij=1}^{\infty} \left[
\frac{1}{2} \hat{B}_i^+ R_{ij} \hat{B}_j^+
+ \hat{B}_i^+       T_{ij}       \hat{B}_j^-       +       \frac{1}{2}
\hat{B}_i^-R_{ij}^*\hat{B}_j^-
\right] + \varepsilon_0.
$$
Consider the linear canonical Bogoliubov transformation, transforming
the initial condition
$B^{\pm}_k(-T)$ for the set of equations
\r{24}-\r{25} to the solution of this set
$B^{\pm}_k(T)$ at $t=T$.
The Wick symbol of the evolution operator
$U^{\Lambda}_{-TT}=:U_{\Lambda}(B^+,B^-):$ for this theory
is presented as \c{5}:
$$
\matrix{
U(B^*,B)=
\frac{\exp(i\int_{-T}^T d\tau [\frac{1}{2}\sum_{i=1}^{\infty} T_{ii}
- \varepsilon_0])}{\sqrt{det G_{\Lambda}}}
\cr
\times
\exp \sum_{ij=1}^{\infty}
\left[-\frac{1}{2} B_i (G_{\Lambda}^{-1}F_{\Lambda}^*)_{ij} B_j
+ B_i   (G_{\Lambda}^{-1}-1)_{ij}   B_j^*    +    \frac{1}{2}    B_i^*
(F_{\Lambda}G_{\Lambda}^{-1})_{ij}
B_j^*
\right]
}
\l{a1}
$$
where
$$
(F_{\Lambda})_{ij} = \frac{\partial B_i^-(T)}{\partial B_j^+(-T)},
(G_{\Lambda})_{ij} = \frac{\partial B_i^+(T)}{\partial B_j^+(-T)}.
$$
According to the appendix, the conditions
$$
\begin{array}{c}
\sum_{ij} |(G_{\Lambda})_{ij} - G_{ij}|^2 \to_{\Lambda\to \infty} 0,
\sum_{ij} |(F_{\Lambda})_{ij} - F_{ij}|^2 \to_{\Lambda\to \infty} 0,\\
det G_{\Lambda} \to_{\Lambda\to \infty} det G
\end{array}
\l{k1}
$$
imply that the operator
$U^{\Lambda}_{-TT}$ has a strong limit
as
$\Lambda\to\infty$. To check condition \r{k1}, it is necessary to show
that

(a) the $l^2$-vectors of the form
$\frac{\partial Q^{(s)}(T)}{\partial
a_k^+(-T)}$ and
$\frac{\partial a_k^+(T)}
{\partial Q^{(s)}(-T)}$ have strong limits as $s=0,...,2l-1$.

(b) the operators with matrices
$\frac{\partial a_k^+(T)}{\partial
a_m^{\pm}(-T)}$ are presented as
$$
\frac{\partial a_k^+(T)}{\partial
a_m^{-}(-T)} =     (A_{\Lambda}^{(1)}A_{\Lambda}^{(2)}
A_{\Lambda}^{(3)})_{km},
\frac{\partial a_k^+(T)}{\partial
a_m^{+}(-T)} =     (1+ A_{\Lambda}^{(4)}A_{\Lambda}^{(5)}
A_{\Lambda}^{(6)})_{km},
$$
for some  operators  $A^{(i)}$  that converge as $\Lambda\to\infty$ in
the norm
$$
||A||_2 = \sqrt {Tr A^+A}.
\l{k8}
$$

Namely, under conditions (a) and (b) the operators
$F_{\Lambda}$ and    $G_{\Lambda}$    are    evidently   converge   as
$\Lambda\to\infty$, while the determinant $det
G_{\Lambda}$ converges because of lemma 2 of the appendix.

The matrices are expressed via the fundamental solution of the equation
$$
\sum_{s=0}^{2l} (z_s     Q_t^{(s)})^{(s)}     +     (-1)^{l+1}     g_t
\int_{-\infty}^t d\tau  (g_{\tau} Q_{\tau})^{(2l)} \sum_{k=1}^{\infty}
\frac{\mu_k^2}{\Omega_k^{2l+1}} \sin[\Omega_k (t-\tau)] = j_t.
\l{k2}
$$
The solution of eq.\r{k2} is expressed via the linear combination
of the initial conditions and the right-hand side:
$$
Q_t = \sum_{s=0}^{2l-1} c_s(t) Q^{(s)}(-T)  +  \int_{-\infty}^t  d\tau
G_{t\tau} j_{\tau}.
$$
Let the function $G_{t\tau}$  be equal to zero
at   $t<\tau$. Then the integration can be supposed to be taken
from $-\infty$ to $+\infty$.

It follows from eqs.\r{27} and \r{28} that
$$
\begin{array}{c}
\frac{\partial Q^{(s)}(T)}{\partial a_k^+(-T)}
= -\int_{-\infty}^{\infty} d\tau
\frac{\partial^s}{\partial T^s}
\frac{\partial^l}{\partial \tau^l}
(G_{T\tau}g_{\tau}) \frac{i^l \mu_k e^{i\Omega_k{\tau}}}
{\sqrt{2}\Omega_k^{l+1/2}},
\\
\frac{\partial a_k^+(T)}{\partial Q^{(s)}(-T)}
= i^{l+1}
\int_{-\infty}^{\infty} d\tau
\frac{\partial^l}{\partial \tau^l}
(g_{\tau}c_s(\tau)) \frac{\mu_k e^{i\Omega_k{\tau}}}
{\sqrt{2}\Omega_k^{l+1/2}},
\end{array}
\l{k6}
$$
$$
\begin{array}{c}
\frac{\partial a_k^+(T)}{\partial a_m^+(-T)}
= \delta_{km}
- i (-1)^l
\int d\tau_1 d\tau_2
 \frac{\mu_k e^{i\Omega_k{\tau}_1}}
{\sqrt{2}\Omega_k^{l+1/2}}
 \frac{\mu_m e^{i\Omega_k\tau_2}}
{\sqrt{2}\Omega_m^{l+1/2}}
\frac{\partial^l}{\partial \tau_1^l}
\frac{\partial^l}{\partial \tau_2^l}
(g_{\tau_1}G_{\tau_1\tau_2}
g_{\tau_2})
\\
\frac{\partial a_k^+(T)}{\partial a_m^-(-T)}
=
- i
\int d\tau_1 d\tau_2
 \frac{\mu_k e^{i\Omega_k{\tau}_1}}
{\sqrt{2}\Omega_k^{l+1/2}}
 \frac{\mu_m e^{-i\Omega_k\tau_2}}
{\sqrt{2}\Omega_m^{l+1/2}}
\frac{\partial^l}{\partial \tau_1^l}
\frac{\partial^l}{\partial \tau_2^l}
(g_{\tau_1}G_{\tau_1\tau_2}
g_{\tau_2})
\end{array}
\l{k7}
$$

To justify the properties (a) and (b),  it is sufficient to prove  the
uniform convergence of the functions
$$
\frac{\partial^s}{\partial T^s}
\frac{\partial^l}{\partial \tau^l}
(G_{T\tau}g_{\tau}),\quad
\frac{\partial^l}{\partial \tau^l}
(g_{\tau}c_s(\tau)),\quad
\frac{\partial^l}{\partial \tau_1^l}
\frac{\partial^l}{\partial \tau_2^l}
(g_{\tau_1}G_{\tau_1\tau_2}
g_{\tau_2})
\l{k4}
$$
at $[-T,T]$.  Namely, the property \r{13} implies the convergence
of vectors \r{k6}.  Construct operators $A^{(i)}$.  The operator
$A^{(3)}$ transforms the sequence
$f_m$ from $l^2$ to the function
$
(A^{(3)}f)(\tau) = \sum_m
 \frac{\mu_m e^{-i\Omega_m{\tau}}}
{\sqrt{2}\Omega_m^{l+1/2}}f_m
$, let $A^{(2)}$ be an integral operator with the kernel
$\frac{\partial^l}{\partial \tau_1^l}
\frac{\partial^l}{\partial \tau_2^l}
(g_{\tau_1}G_{\tau_1\tau_2}
g_{\tau_2})$, while the operator
$A^{(1)}$  transforms the function
$\varphi$ from
$L^2[-T,T]$ to the sequence
$
(A^{(1)}\varphi)_m =
-i \int d\tau
\frac{\mu_m e^{i\Omega_m{\tau}}}
{\sqrt{2}\Omega_m^{l+1/2}}f_m
$.  Choose $A^{(5)}=A^{(2)}$, $A^{(4)}=A^{(1)}$,
$ (A^{(6)}f)(\tau) = \sum_m
 \frac{(-1)^l \mu_m e^{i\Omega_m{\tau}}}
{\sqrt{2}\Omega_m^{l+1/2}}f_m
$.
Convergence of these operator in the \r{k8}-norm is a corollary of the
property \r{13} and convergence of the functions \r{k4}.

Convergence of the function
$
\frac{\partial^l}{\partial \tau^l}
(g_{\tau}c_s(\tau))$ is a corollary
of the lemma 3 of the appendix.
To investigate the property of convergence of the function
$G^{(l)(l)} \equiv
\frac{\partial^l}{\partial \tau_1^l}
\frac{\partial^l}{\partial \tau_2^l}
G_{\tau_1\tau_2}
$,
represent eq.\r{k2} in the form
$$
\begin{array}{c}
\sum_{s=0}^{l} z_s
\left(\frac{d}{dt}\right)^{2s-2l}
G^{(l)(l)}_{tt_0} + (-1)^{l+1}
\left(\frac{d}{dt}\right)^{-l} g
\left(\frac{d}{dt}\right)^l \hat{K}
\left(\frac{d}{dt}\right)^l
g
\left(\frac{d}{dt}\right)^{-l} G^{(l)(l)}_{tt_0}
\\
= (-1)^l \delta(t-t_0),
\end{array}
$$
where
$\left(\frac{d}{dt}\right)^{-1}$ is an integral operator
\\
$
(\left(\frac{d}{dt}\right)^{-1} f)(t)   =  \int  _{-\infty}^t  f(\tau)
d\tau$, while $\hat{K}$ is the operator with the kernel
$$
K(t,\tau) = \sum_{k=1}^{\infty} \frac{\mu_k^2}{\Omega_k^{2l+1}}
\sin [\Omega_k (t-\tau)].
$$
Convergence of functions
$G^{(l)(l)}$ and
$\frac{\partial^m}{\partial t^m}
\frac{\partial^n}{\partial \tau^n}
G_{t\tau}$, $m,n\le l$,
is a corollary of lemma 3.

As $t>T>\tau$, the functions $G_{t\tau}$ obey the equation
$
\sum_{s=0}^{2l} z_s
\frac{\partial^s}{\partial t^{2s}}
G_{t\tau} =0
$.
Therefore
$$
\matrix{
G_{tt_0} =
\sum_{s=0}^{2l-1}
\frac{\partial^s}{\partial T^{s}}
G_{Tt_0} \sum_{m= [\frac{s}{2}+1]}^l
\frac{(t-T)^{2l-2m+s}}{(2l-2m+s)!}\frac{z_m}{z_l}
\cr
- \sum_{m=0}^{l-1} \int_T^t d\tau G_{\tau t_0}
\frac{z_m}{z_l} \frac{(t-\tau)^{2l-1-2m}}{(2l-1-2m)!}
}
\l{31}
$$
The quantity
$\frac{\partial^s}{\partial T^{s}}
G_{Tt_0}$ is expressed
via linear combinations
of the values
$G_{\tau t_0}$ at $\tau
\in (T,T+\delta)$ from eq.\r{31} for
$t=t_1,...,t_{2s} \in (T,T+\delta)$.
Therefore, the convergence of functions
$\frac{\partial^s}{\partial T^s}
\frac{\partial^m}{\partial \tau^m}
G_{T\tau}$ is a corollary of convergence of the quantity
$\frac{\partial^m}{\partial \tau^m}
G_{T\tau}$.

Thus, the strong convergence of the Bogoliubov $S$-matrix as
$\Lambda\to\infty$ is checked for
smooth functions
$g$ which vanish for $t \notin (-T,T)$.
Convergence of the operator $S^+$ is checked analogously.

\subsection{Representations of different operators}

In this  subsection  we  investigate  what   operators   $O_{\Lambda}$
transform  the  vectors of the type $T_{\Lambda}\Psi_1$ to the vectors
of the same type $T_{\Lambda}\Psi_2$.  This means that  the  condition
\r{16}  is invariant under transformation $O_{\Lambda}$.  In this case
the operator \r{b4} is regular as $\Lambda\to\infty$.

Consider the Heisenberg operators
$O_{\Lambda}=\hat{a}_k^+(t)=e^{iHt}\hat{a}_k^+e^{-iHt}$.
Since the operators
$e^{-iHt}$ and $T_{\Lambda}$ can be expressed
via the evolution operator
$V_{t_1t_2}$ for the theory with the Hamiltonian
\r{23},
$$
e^{-iHt}=e^{-iH_qt}U_{0t}, \qquad T_{\Lambda}=U_{-T 0} e^{iH_QT}
$$
for Heisenberg operators
$\hat{a}_k^{+}(t)$ in the representation
\r{b4} one has
$$
T_{\Lambda}^+\hat{a}_k^+(t)T_{\Lambda} =
e^{-iH_QT} U^+_{-Tt} e^{iH_qt} \hat{a}_k^+
e^{-iH_qt}U_{-Tt} e^{iH_QT}
$$
$$
= e^{-iH_QT} U^+_{-Tt} \hat{a}_k^+ U_{-Tt} e^{iH_QT} e^{i\Omega_kt}.
$$
Analogously, one has
$$
T_{\Lambda}^+\hat{a}_k^-(t)T_{\Lambda} =
 e^{-iH_QT} U^+_{-Tt} \hat{a}_k^- U_{-Tt} e^{iH_QT} e^{-i\Omega_kt}.
$$
$$
T_{\Lambda}^+\hat{Q}(t)T_{\Lambda} =
 e^{-iH_QT} U^+_{-Tt} \hat{Q} U_{-Tt} e^{iH_QT} e^{-i\Omega_kt}.
$$
To investigate   the   regularity  of  the  operators  \r{b4},  it  is
sufficient to investigate the regularity as
$\Lambda\to\infty$ of the operators
$$
V^+_{-Tt} \hat{a}_k^{\pm} V_{-Tt}, \qquad
V^+_{-Tt} \hat{Q} V_{-Tt}.
\l{b23}
$$
The Heisenberg  equations of motion for the operators \r{b23} coincide
with the classical equations  in  the  model  \r{23}.  Therefore,  for
operators \r{b23} one has
$$
U^+_{-Tt} \hat{a}_k^{\pm} U_{-Tt} =
\sum_{m=1}^{\infty}
\left(
\frac{\partial a_k^{\pm}(t)}{\partial a_m^-(-T)} \hat{a}_m^-
+ \frac{\partial a_k^{\pm}(t)}{\partial a_m^+(-T)} \hat{a}_m^+
\right)
+ \sum_{s=0}^{2l-1}
\frac{\partial a_k^{\pm}(t)}{\partial Q^{(s)}(-T)} \hat{Q}^{(s)}
$$
$$
\l{b24}
$$
$$
U^+_{-Tt} \hat{Q} U_{-Tt} =
\sum_{m=1}^{\infty}
\left(
\frac{\partial Q(t)}{\partial a_m^-(-T)} \hat{a}_m^-
+ \frac{\partial Q(t)}{\partial a_m^+(-T)} \hat{a}_m^+
\right)
+ \sum_{s=0}^{2l-1}
\frac{\partial Q(t)}{\partial Q^{(s)}(-T)} \hat{Q}^{(s)}.
$$
It follows from the previous subsection that the operators
$T_{\Lambda}^+a_k^{\pm}T_{\Lambda}$ and
$T_{\Lambda}^+Q^{(s)}T_{\Lambda}$, $s=\overline{0,l}$ are regular as
$\Lambda\to\infty$. Equations of motion imply that the operators
$T_{\Lambda}^+P_sT_{\Lambda}$, $s=\overline{0,l-1}$, are singular
as $\Lambda\to\infty$.

\section{Non-unitarity of evolution in the external field}

It will   be  shown  in  this  section  that  the  evolution  operator
corresponding to the model \r{a1} in the external field in the leading
order of perturbation theory may be nonunitary,  since it is necessary
to consider different representations  of  the  canonical  commutation
relations at different moments of time.

The quantum theory in the external field  is  usually  constructed  as
follows \c{4}. The field
$Q(t)$ is decomposed into two parts. The ''classical''
part $\frac{1}{g}Q_c(t)$ is of order $O(1/g)$. The remaining part
$\hat{X}(t)$ is ''quantum'',
$$
Q(t) = \frac{1}{g}Q_c(t) + \hat{X}(t).
\l{a1*}
$$
The classical part of the field
$q_k$ will be set to zero.
Action \r{a1} takes the following form:
$$
S= const - \frac{1}{g} \int dt \hat{X} \sum_{s=0}^l z_s Q_c^{(2s)}
+ \int dt [L_X + \sum_{k=1}^{\infty}
\left(\frac{\dot{q}_k^2}{2} - \frac{\Omega_k^2q_k^2}{2}\right)
- \sum_{k=1}^{\infty} \mu_k q_k Q_c] + O(g).
$$
The term of order
$1/g$  vanishes if the ``external field''
$Q_c(t)$ obeys the classical equation of motion
$$
\sum_{s=0}^l z_s Q_c^{(2s)} = 0,
$$
which can be obtained from eq.\r{a1}
by the variation procedure as
$g\to 0$.
Neglect the terms of order
$O(g)$.
We obtain that the degrees of freedom corresponding to fields
$\hat{X}$ and $q_k$ are independent.
Thus, one can consider the problem of quantization of the fields
$q_k$ in the external nonstationary classical field
$Q_c(t)$.  The Hamiltonian of this model has the form:
$$
H = \sum_{k=1}^{\infty}
\left(\frac{{p}_k^2}{2} + \frac{\Omega_k^2q_k^2}{2}\right)
+ \sum_{k=1}^{\infty} \mu_k q_k Q_c(t).
\l{a2i}
$$

\subsection{Fock representation: range of validity}

One can  try  to  use  the  Fock  representation  of   the   canonical
commutation relations  \r{a2*}.  Let us investigate in what case it is
possible.

Under this choice of the representation,  the Hamiltonian \r{a2i} takes
the form:
$$
H=\sum_k \Omega_k     a_k^+a_k^-     +    \sum_{k=1}^{\infty}    \mu_k
\frac{a_k^++a_k^-}{\sqrt{2\Omega_k}}Q_c + E_0
\l{a3*}
$$
If we choose the Wick ordering of creation and annihilation operators,
the constant
$E_0$ vanishes.

Consider the solution to the Schrodinger equation
$$
i\frac{d\Psi}{dt} = H\Psi,
\l{a4}
$$
which has the form of the coherent state
$$
\Psi(t)= c(t)\exp[\sum_{k=1}^{\infty} \alpha_k(t)\hat{a}_k^+] |0>,
\l{a5i}
$$
being expressible via the vacuum vector
$|0>$  of the form  $(1,0,0,...)$  and complex functions
$c(t)$ and $\alpha_k(t)$. Substitution of the vector
\r{a5i} to eq.\r{a4} leads us to the relations
$$
i\dot{c} = \left[ \sum_{k=1}^{\infty}
\frac{\mu_k\alpha_k}{\sqrt{2\Omega_k}}{\cal Q} + E_0\right]c;
\qquad
i\dot{\alpha_k} = \Omega_k \alpha_k + \frac{\mu_k}{\sqrt{2\Omega_k}}
{\cal Q}.
$$
The divergences appearing in the multiplier
$c$ can be eliminated by the proper choice of the
''counterterm'' $E_0$. Investigate now the functions
$\alpha_k(t)$:
$$
\alpha_k(t) = \alpha_k(0)e^{-i\Omega_k t} + \rho_k(t),
\l{a6}
$$
where
$$
\rho_k(t) = - i\frac{\mu_k}{\sqrt{2\Omega_k}}
\int_0^t d\tau {\cal Q}(\tau) e^{-i\Omega_k (t-\tau)}.
\l{7}
$$
It follows from  \c{5}
that expression \r{a5i} defines the Fock vector
if
$$
\sum_k |\alpha_k(t)|^2 < \infty.
\l{8}
$$
Integrating eq.\r{7} by parts, we obtain
$$
\rho_k(t) = \beta_k^l(t) - \beta_k^l(0)e^{-i\Omega_kt} + \gamma_k^l(t),
\l{9a}
$$
where
$$
\matrix{
\beta_k^l(t) = - \frac{\mu_k}{\sqrt{2\Omega_k}} \sum_{s=0}^{l-2}
i^s {\cal Q}^{(s)} (t) \frac{1}{\Omega_k^{s+1}};
\cr
\gamma_k^l(t) =  -\frac{\mu_k}{\sqrt{2\Omega_k}}  \int_0^t  i^l  d\tau
{\cal Q}^{(l-1)}(\tau) \frac{e^{i\Omega_k(\tau-t)}}{\Omega_k^{l-1}}.
}
$$
The leading in
$1/\Omega_k$ order is
$
\rho_k(t) \simeq -\frac{\mu_k}{\sqrt{2\Omega_k^3}}
({\cal Q}(t)-{\cal Q}(0)e^{-i\Omega_kt})
$.
Condition   \r{8} is satisfied if
$$
\sum_{k=1}^{\infty} \frac{\mu_k^2}{\Omega_k^3} < \infty.
\l{9}
$$
Thus, the evolution transformation in the external field
can be viewed as  an  unitary  operator  if  the  condition  \r{9}  is
satisfied.

Condition  \r{9} can be also obtained as follows.
Heisenberg equations of motion for the operators
$\pi(a_k^{\pm}(t))
= e^{i\hat{H}t}\pi(a_k^{\pm}) e^{-i\hat{H}t}$ are written as
$$
\mp i \frac{d}{dt}
\pi({a}_k^{\pm}(t)) = \Omega_k \pi(a_k^{\pm}(t)) +
\frac{\mu_k}{\sqrt{2\Omega_k}} {\cal Q}(t).
\l{10}
$$
Heisenberg creation  and  annihilation  operators  at  different  time
moments are related as
$$
\pi(a_k^{\pm}(t)) = e^{-i\Omega_kt}\pi(a_k^{\pm}(0)) + \rho_k^{\pm}(t),
\l{11}
$$
where $\rho_k^-=\rho_k$, $\rho_k^+=\rho^*_k$,
$\pi(a_k^{\pm}(0))=\hat{a}_k^{\pm}$.
According to  \c{5}, the canonical transformation
\r{11} is unitary
if and only if  $\sum_k
|\rho_k(t)|^2<\infty$. This condition is equivalent to \r{9}.

\subsection{Different Hilbert spaces at different time moments}

If the condition \r{9} is not satisfied,  one should consider non-Fock
representations of CCR in order to construct the quantum theory. Since
the choice   of  the  non-Fock  representation  is  specified  by  the
interaction (see,  for example,  \c{12}), which depends on time in our
case, it  is necessary to consider different representations of CCR at
different time moments.

Consider this hypothesis for the model \r{a3*}.

One can consider the ''large'' linear state space $\cal L$ and specify
the subspaces  ${\cal H}_{\alpha}\subset {\cal L}$.  The inner product
is introduced on each  subspace  ${\cal  H}_{\alpha}$.  The  parameter
$\alpha$ belongs to some set $A$. The operators $e^{i\sum_k \hat{a}_k^+
z_k}$ defined on $\cal L$ transform elements of ${\cal H}_{\alpha}$ to
elements of   ${\cal   H}_{\alpha}$.   The  restrictions  $\pi_{\alpha}
(a_k^{\pm})= \hat{a}_k^{\pm}|_{{\cal  H}_{\alpha}}$  of  creation  and
annihilation operators  on  the subspace ${\cal H}_{\alpha}$ specifies
the $\alpha$-representation of CCR.

The evolution operator $U_t$ is defined as a set of mappings $u_t:A\to
A$ and $V_t^{\alpha}:  {\cal H}_{\alpha} \to {\cal H}_{u_t\alpha}$. If
the initial condition $\Psi_0$ belongs  to  ${\cal  H}_{\alpha}$,  the
state at  time  moment  $t$  is  defined  as  $V_t^{\alpha}\Psi_0$ and
belongs to ${\cal H}_{u_t\alpha}$.

Choose as  a  space  $\cal  L$  the  space  of  analytic   functionals
$\Psi(z)=\Psi(z_1,z_2,...)$. The  creation  and annihilation operators
have the form:
$$
\hat{a}_k^+ = z_k ,\qquad \hat{a}_k^- = \frac{\partial }{\partial z_k}.
\l{j1}
$$
Define the subset ${\cal H}_0 \in {\cal L}$ as follows.  Consider  the
expansion of the functional $\Psi$ into a series:
$$
\Psi =    \sum_{n=0}^{\infty}   \frac{1}{\sqrt{n!}}   \sum_{i_1...i_n}
\Psi^{(n)}_{i_1...i_n} z_{i_1}...z_{i_n}
$$
with symmetric     in      $i_1...i_n$      coefficient      functions
$\Psi^{(n)}_{i_1...i_n}$. We say that $\Psi\in {\cal H}_0$ if
$$
||\Psi||^2 =           \sum_{n=0}^{\infty}            \sum_{i_1...i_n}
|\Psi^{(n)}_{i_1...i_n}|^2 < \infty.
\l{j2}
$$
Introduce the inner product on ${\cal H}_0$:
$$
(\tilde{\Psi},{\Psi})_0 =                    \sum_{n=0}^{\infty}
\tilde{\Psi}^{(n)*}_{i_1...i_n} \Psi^{(n)}_{i_1...i_n}.
$$
Note that the space ${\cal H}_0$ is isomorphic to the Fock space \c{5}.

Let $\alpha=(\alpha_1,\alpha_2,...)$  is a set of complex umbers.  Say
that $\Psi \in {\cal H}_{\alpha}$ if the functional
$$
w_{\alpha} \Psi     (z)     =     e^{-\sum_{k=1}^{\infty}     \alpha_k
(z_k+\alpha_k^*)} \Psi (z+\alpha^*)
$$
belongs to  ${\cal  H}_0$.  Introduce on ${\cal H}_{\alpha}$ the inner
product:
$$
(\tilde{\Psi},{\Psi})_{\alpha}                                       =
(w_{\alpha}\tilde{\Psi},w_{\alpha}\Psi)_0
$$
In particular,   the   functional   $\Psi(z)=   e^{\sum_{k=1}^{\infty}
z_k\alpha_k}$ belongs  to  ${\cal  H}_{\alpha}$  in  any  case.   This
functional belongs to ${\cal H}_0$ if and only if $\alpha\in l^2$.

Let $\Psi_0\in   {\cal   H}_{\alpha(0)}$.  Define  the  mapping  $u_t:
\alpha(0) \mapsto \alpha(t)$ according to \r{a6}.  Since the evolution
equation in the representation \r{j1} has the form
$$
i\dot{\Psi}^t(z) = \left(
\sum_{k=1}^{\infty} \Omega_k  z_k  \frac{\partial  }{\partial  z_k}  +
\sum_{k=1}^{\infty}
\mu_k \frac{z_k+
  \frac{\partial  }{\partial  z_k} }{\sqrt{2\Omega_k}}Q_c + E_0
\right)\Psi^t (z).
$$
The functional $\Phi^t = w_{\alpha(t)} \Psi^t$ obeys the equation:
$$
i\dot{\Phi}^t(z) =
\left(
\sum_{k=1}^{\infty} \Omega_k  z_k  \frac{\partial  }{\partial  z_k}  +
\sum_{k=1}^{\infty}
\mu_k \frac{\alpha_k}{\sqrt{2\Omega_k}}Q_c + E_0
\right)\Phi^t (z).
$$
If $\Phi^0    \in    {\cal   H}_0$,   one   has   $\Phi^t(z)   =   b^t
\Phi^0(ze^{-i\Omega t}))$  for  some  multiplier  $c^t$.   Under   the
appropriate choice  of the counterterm $E_0$,  $\Phi^t\in {\cal H}_0$.
Thus, $\Psi^t \in {\cal H}_{\alpha(t)}$.

\subsection{An algebraic approach}

The lack of the approach of the previous  subsection  is  that  it  is
necessary to  eliminate  the  divergences from the multiplier $b^t$ by
renormalization of $E_0$. If one considered the density matrix instead
of the   wave   function,   this   difficulty   does  not  arise.  The
generalization of the density-matrix approach to systems  of  infinite
number of  degrees  of  freedom  is  the  algebraic  approach \c{12}
which is
suitable for the case when different representations of CCR  arise  at
different time moments.

\subsubsection{Density matrix   for  systems  of  infinite  number  of
degrees of freedom}

In the $d$-dimensional quantum mechanics,  one can use  not  only  the
wave-function language but also the density-matrix conception.  If the
system is  in  a  pure  state  with  the  wave  function  $\Psi$,  the
Blokhintsev-Wigner density matrix is determined as:
$$
\rho(p,q) =      \frac{1}{(2\pi)^d}      \int      d\xi      e^{ip\xi}
\Psi(q-\frac{\xi}{2}) \Psi^*(q+ \frac{\xi}{2}).
\l{v1}
$$
A remarkable property of the density \r{v1} is that the average values
of observables $\hat{A}= A(\hat{p},\hat{q})$ can  be  presented  in  a
form analogous to the classical statistical mechanics:
$$
<\hat{A}> = (\Psi, \hat{A}\Psi) = \int dpdq A(p,q) \rho(p,q),
$$
provided that  the  Weyl  ordering  of the coordinate and momenta
operators are chosen.  For pure states,  the density matrix  specifies
the wave function $\Psi$ up to a multiplier.

For the  case  of infinite number of degrees of freedom,  the numerous
difficulties with  the  divergences  arise.  Nevertheless,   one   can
consider the Fourier transformation of $\rho$:
$$
\tilde{\rho}(\alpha,\beta) =   \int  dp  dq  \rho(p,q)  e^{-i\alpha  p
-i\beta q}.
$$
For pure states \r{v1}, it can be presented as
$$
\hat{\rho}(\alpha,\beta) =      (\Psi,       e^{-i\alpha\hat{p}-i\beta
\hat{q}}\Psi).
\l{v2}
$$
The function \r{v2} can be used  instead  of  the  density  matrix  in
calculations of the average values of observables.

The advantage  of  using  the  function  \r{v2}  is the possibility of
generalization to the infinite-dimensional case.  One can specify  the
state of the system by the average values
$$
\tilde{\rho}(z,z^*) =  <e^{\sum_k (z_k \hat{a}_k^+ - z_k^* \hat{a}_k^-
)}>.
\l{v3}
$$

Consider some examples of ''densities'' \r{v3}.

1. For the vacuum state
$$
\tilde{\rho}(z,z^*) = (\Phi^{(0)}
e^{\sum_k (z_k \hat{a}_k^+ - z_k^* \hat{a}_k^- )} \Phi^{(0)})
= e^{-\frac{1}{2}\sum_k z_k^* z_k}.
$$

2. For the coherent state
$$
\Phi =
e^{\sum_k (\alpha_k \hat{a}_k^+ - \alpha_k^* \hat{a}_k^-)}
\Phi^{(0)},
$$
where $\alpha\in l^2$, one has
$$
\tilde{\rho}(z,z^*) =
e^{\sum_k (\alpha_k^* z_k - \alpha_k z_k^* - \frac{1}{2}z_k^*z_k)}.
\l{v4}
$$

3. Suppose that the non-Fock representation $\pi(a_l^{\pm})$ of CCR in
the space  $\cal  H$  is chosen.  For this case,  one can also use the
''density'' \r{v3}:
$$
\tilde{\rho} (z,z^*) =
(\Phi
e^{\sum_k (z_k \pi({a}_k^+) - z_k^* \pi({a}_k^-) )} \Phi),
\qquad \Phi \in {\cal H}.
\l{v5}
$$

4. As an example,  consider the following ''$\alpha$- representation''
of CCR:
$$
{\cal H} ={\cal F}, \qquad
\pi_{\alpha}(a_k^+) = \hat{a}_k^+ +\alpha_k^*, \qquad
\pi_{\alpha}(a_k^-) = \hat{a}_k^- +\alpha_k.
\l{v6}
$$
For the vacuum state $\Phi=\Phi^{(0)}$, the ''density'' $\tilde{\rho}$
has the form \r{v4}, but the case $\alpha\in l^2$ can be involved.

{\it Definition}.  We say  that  the  function  $\rho(z,z^*)$  is  an
{\it $\alpha$-density}  if  it  is  written  in the form \r{v5} for the
representation \r{v6}.  For  $\alpha=0$,  $\alpha$-densities  will  be
called as Fock densities.

{\it Statement 1}. $\tilde{\rho}(z,z^*)$ is an $\alpha$-density if and
only if
$\tilde{\rho}(z,z^*) e^{\sum_{k=1}^{\infty} (-\alpha_k^*z_k +  \alpha_k
z_k^*)}$  is a Fock density.

{\it Proof}.   The   fact    that    $\tilde{\rho}(z,z^*)$    is    an
$\alpha$-density means that
$$
\tilde{\rho}(z,z^*)=
(\Phi
e^{\sum_k (z_k (\hat{a}_k^+ +\alpha_k^*)
- z_k^* (\hat{a}_k^-+\alpha_k) )} \Phi),
$$
for some vector $\Phi  \in  {\cal  F}$.  This  is  equivalent  to  the
statement that
$\tilde{\rho}(z,z^*)
e^{\sum_{k=1}^{\infty} (-\alpha_k^*z_k +  \alpha_k z_k^*)}$
is a Fock density.

{\it Statement 2}. Let $\tilde{\rho}$ be an $\alpha^{(1)}$-density.
Then
$\tilde{\rho}$ is an $\alpha^{(2)}$-density if and only if
$\alpha^{(2)} - \alpha^{(1)} \in l^2$.

{\it Proof}. According to statement 1, the function
$$
\tilde{\rho}(z,z^*)
e^{\sum_{k=1}^{\infty} (-\alpha_k^{(1)*}z_k +  \alpha_k^{(1)} z_k^*)}
= f(z,z^*) =
(\Phi, e^{\sum_k (z_k \hat{a}_k^+ - z_k^* \hat{a}_k^- )} \Phi)
\l{yy1}
$$
is a Fock density. Let
$\alpha^{(2)} - \alpha^{(1)} \in l^2$ and
$$
\Phi_1 =
e^{\sum_k ((\alpha_k^{(1)}- \alpha_k^{(2)})
\hat{a}_k^+ -
((\alpha_k^{(1)}- \alpha_k^{(2)}))^*
\hat{a}_k^- )} \Phi.
$$
One has
$$
\tilde{\rho}(z,z^*)
e^{\sum_{k=1}^{\infty} (-\alpha_k^{(2)*}z_k +  \alpha_k^{(2)} z_k^*)}
=
(\Phi_1, e^{\sum_k (z_k \hat{a}_k^+ - z_k^* \hat{a}_k^- )} \Phi_1).
\l{y2}
$$
Therefore, $\tilde{\rho}$ is an $\alpha^{(2)}$-density.

Let $\tilde{\rho}$ be an $\alpha^{(2)}$-density, $z\in l^2$ and
$$
z_k^{(n)}=z_k , k\le n, \qquad z_k^{(n)}=0, k>n.
$$
It follows from eqs.\r{yy1} and \r{y2} that
$$
\frac{
(\Phi_1,
e^{{\varepsilon} \sum_k    (z_k^{(n)}    \hat{a}_k^+    -   z_k^{(n)*}
\hat{a}_k^- )} \Phi_1)
}
{
(\Phi,
e^{{\varepsilon} \sum_k    (z_k^{(n)}    \hat{a}_k^+    -   z_k^{(n)*}
\hat{a}_k^- )} \Phi)
}
=
e^{{\varepsilon}\sum_k (
(\alpha_k^{(2)}-\alpha_k^{(1)})z_k^{(n)*}
-
(\alpha_k^{(2)*}-\alpha_k^{(1)*})z_k^{(n)})
}.
\l{y3}
$$
It follows from the corollary 3 of lemma 1 from the Appendix that
$$
\begin{array}{c}
(\Phi_1,
e^{{\varepsilon} \sum_k    (z_k^{(n)}    \hat{a}_k^+    -   z_k^{(n)*}
\hat{a}_k^- )} \Phi_1) \to_{n\to\infty}
(\Phi_1,
e^{{\varepsilon} \sum_k    (z_k    \hat{a}_k^+    -   z_k^{*}
\hat{a}_k^- )} \Phi_1)
\\
(\Phi,
e^{{\varepsilon} \sum_k    (z_k^{(n)}    \hat{a}_k^+    -   z_k^{(n)*}
\hat{a}_k^- )} \Phi) \to_{n\to\infty}
(\Phi,
e^{{\varepsilon} \sum_k    (z_k    \hat{a}_k^+    -   z_k^{*}
\hat{a}_k^- )} \Phi)
\end{array}
$$
Corollary 3 also implies that for sufficiently small ${\varepsilon}$
$$
\begin{array}{c}
(\Phi_1,
e^{{\varepsilon} \sum_k    (z_k    \hat{a}_k^+    -   z_k^{*}
\hat{a}_k^- )} \Phi_1) \ne 0,
\\
(\Phi,
e^{{\varepsilon} \sum_k    (z_k    \hat{a}_k^+    -   z_k^{*}
\hat{a}_k^- )} \Phi)\ne 0.
\end{array}
$$
This implies that
$$
\begin{array}{c}
(\Phi_1,
e^{{\varepsilon} \sum_k    (z_k^{(n)}    \hat{a}_k^+    -   z_k^{(n)*}
\hat{a}_k^- )} \Phi_1) \ne 0,
\\
(\Phi,
e^{{\varepsilon} \sum_k    (z_k^{(n)}    \hat{a}_k^+    -   z_k^{(n)*}
\hat{a}_k^- )} \Phi) \ne 0
\end{array}
$$
at $n\ge n_1$. Therefore, the eft-hand side of eq.\r{y3} tends to
$$
\frac{
(\Phi_1,
e^{{\varepsilon} \sum_k    (z_k^{}    \hat{a}_k^+    -   z_k^{*}
\hat{a}_k^- )} \Phi_1)
}
{
(\Phi,
e^{{\varepsilon} \sum_k    (z_k^{}    \hat{a}_k^+    -   z_k^{*}
\hat{a}_k^- )} \Phi)
}.
$$
This means that the limit
$$
\lim_{n\to\infty}
\sum_k (
(\alpha_k^{(2)}-\alpha_k^{(1)})z_k^{(n)*}
-
(\alpha_k^{(2)*}-\alpha_k^{(1)*})z_k^{(n)})
$$
exists for all $z\in l^2$.  Choose $z =  i(\alpha^{(2)}-\alpha^{(1)})$.
We see that $(\alpha^{(2)}-\alpha^{(1)} \in l^2$.
Statement is proved.

We see that the notion of ''density'' \r{v3} is  useful  in  order  to
specify states corresponding to different representations of CCR.  One
can even investigate the case when the representation is time-dependent.

\subsubsection{Evolution of density matrix}

Let us write down the evolution equation for the  average  \r{v3}  for
the system \r{a3*}.

From eq.\r{a4} one has
$$
i\dot{\tilde{rho}}
=  < [e^{\sum_k (z_k \hat{a}_k^+ - z_k^* \hat{a}_k^- )}; H ]>.
\l{v7}
$$
for the  Fock  density  case.  Eq.  \r{v7}  can  be postulates for the
general case as well. It follows from CCR that
$$
f(a^+,a^-)
e^{\sum_k (z_k \hat{a}_k^+ - z_k^* \hat{a}_k^- )}
=
e^{\sum_k (z_k \hat{a}_k^+ - z_k^* \hat{a}_k^- )}
f(a^++\alpha^*,a^-+\alpha).
$$
This implies that
$$
[H;
e^{\sum_k (z_k \hat{a}_k^+ - z_k^* \hat{a}_k^- )}]
=
e^{\sum_k (z_k \hat{a}_k^+ - z_k^* \hat{a}_k^- )}
\left[
\Omega_k z_k^*\hat{a}_k^- + \Omega_k z_k \hat{a}_k^+
+ \Omega_k      z_k^*      z_k      +       \mu_k       {\cal       Q}
\frac{z_k^*+z_k}{\sqrt{2\Omega_k}}
\right].
$$
Furthermore, it follows from CCR that
$$
\frac{\partial}{\partial z_i}
<e^{\sum_k (z_k \hat{a}_k^+ - z_k^* \hat{a}_k^- )}>
=
<e^{\sum_k (z_k \hat{a}_k^+ - z_k^* \hat{a}_k^- )}
(a_i^++ \frac{1}{2}z_i^*)>;
$$
$$
-\frac{\partial}{\partial z^*_i}
<e^{\sum_k (z_k \hat{a}_k^+ - z_k^* \hat{a}_k^- )}> =
<e^{\sum_k (z_k \hat{a}_k^+ - z_k^* \hat{a}_k^- )}
(a_i^-+ \frac{1}{2}z_i^*)>;
$$
Substituting these  relations  to  the  commutator,  we   obtain   the
following evolution equation:
$$
-i\dot{\tilde{\rho}} =
\sum_k\left(
-\Omega_k z_k^* \frac{\partial}{\partial z_k^*}
+ \Omega_k z_k \frac{\partial}{\partial z_k }
+ \mu_k {\cal Q} \frac{z_k+ z_k^*}{\sqrt{2\Omega_k}}
\right)
\tilde{\rho}.
$$
Substitution
$$
\tilde{\rho}_t(z,z^*) =
e^{\sum_{k=1}^{\infty} (\alpha^*_k(t)z_k - \alpha_k(t)z_k^*)}
f(z,z^*)
$$
gives us  eq.\r{a6}  on  the  function $\alpha_k(t)$ and the following
equation on $f$:
$$
-i\dot{f} = \left(
-\Omega_k z^*_k \frac{\partial}{\partial z_k^*}
+\Omega_k z_k \frac{\partial}{\partial z_k}
\right)f.
\l{v8}
$$
{\it Statement 3.} Let  $\tilde{\rho}_0$  be  an  $\alpha(0)$-density.
Then $\tilde{\rho}_t$ is an $\alpha(t)$-density,  where $\alpha(t)$ is
given by eq.\r{a6}.

{\it Proof.} It is sufficient to check that the property that $f$ is a
Fock density is invariant under evolution \r{v8}. Let
$$
f_t = (\Phi_t,
e^{\sum_{k=1}^{\infty} (\hat{a}^+_kz_k - \hat{a}^-_kz_k^*)},
\Phi_t)
$$
where $\Phi_t = e^{-\sum_k \Omega_k \hat{a}_k^+ \hat{a}_k^-}\Phi_0$.
Note that $f_t$ obeys eq.\r{v8}. Statement is proved.

\subsubsection{Time evolution of the representation}

According to statement  2,  the  function  $\alpha_k$  specifying  the
choice of  the  representation is defined up to an element from $l^2$.
We say that $\alpha \sim \alpha'$ if $\alpha-\alpha' \in  l^2$.  Denote
the class of equivalence as $[\alpha]$.

Let the condition \r{13}
be satisfied.  In this case the quantity $\gamma_k^l(t)$  entering  to
eq.\r{9a} is an element of $l^2$, so that
$$
[\alpha_k(t)] =            [(\alpha_k(0)-\beta_k^l(0))e^{-i\Omega_kt}+
\beta_k^l(t)],
$$
where quantities $\beta_k^l(t)$ are expressed via ${\cal Q}(t)$,  ...,
${\cal Q}^{(l-2)}(t)$.  We see that in general case the representation
of CCR at time moment $t$ depends not only  on  the  value  of  ${\cal
Q}^{(s)}(t)$ but also on values ${\cal Q}^{(s)}(0)$. However, there is
a {\it special}  case  when  the  representation  depends  only  on   the
derivatives of  $\cal  Q$  at  the  same  time  moment.  Such  a  case
corresponds to the following choice of the initial representation:
$$
\alpha_k(0) - \beta_k^l(0) \in l^2.
$$
This implies that
$$
\alpha_k(t) - \beta_k^l(t) \in l^2,
$$
so that
$$
[\alpha_k(t)] = [\beta_k^l(t)].
\l{z1}
$$
If the condition \r{13} is satisfied at $l=2$,  the formula \r{z1} for
the representation takes the form:
$$
[\alpha_k(t)] = - \frac{\mu_k}{\sqrt{2\Omega_k^3}} {\cal Q}(t).
\l{z2}
$$
This choice of the representation is in agreement with papers \c{2,3}.
In these   articles   the   processes    in    strong    nonstationary
electromagnetic and  gravitational  fields  were  considered.  It  was
suggested to   consider   the   representation   obtained    by    the
diagonalization procedure of the Hamiltonian at each time moment.

We see  also  that  if the condition \r{13} is not satisfied at $l=2$,
the prescription \r{z2} of \c{2,3} is not valid.

\section{Explanation of the paradox}

We have constructed
in section 2 the renormalized Hilbert state space
and the renormalized evolution operator for  the  model
\r{1} which is unitary.  This implies that it is sufficient to use one
representation of  CCR.  This  is  in  agreement  with  the  Whightman
axiomatic approach.  However,  section 3 tells us that the model \r{1}
in the strong external field in the leading order in $g$  is  unusual:
one should  choose  different representations of CCR at different time
moments. Let us discuss this paradox.

\subsection{
Extraction of the $c$-number from the field in the cutoffed theory}

The procedure  of extracting the $c$-number component of the field $Q$
is justified within the Hamiltonian framework as follows \c{19,20,21}.
Consider the  regularized  theory  with fixed $\Lambda$.  At the fixed
moment of time the state of the
system is specified by sets
${\cal P} = ({\cal P}_0,...,{\cal P}_{l-1})$ and
${\cal Q} = ({\cal Q},...,{\cal Q}^{(l-1)})$ and  regular as $g\to 0$
vector $X$ of the Fock space which can be identified with the state in
the external   field   $\cal  Q$.  The  solution  of  the  regularized
Schrodinger equation \r{15} depends on the small parameter $g$ as
$$
\Psi_t = e^{\frac{i}{g^2} S_t} U_g[{\cal P}_t,{\cal Q}_t]X_t
\l{33}
$$
where the index $\Lambda$ is omitted,
$$
U_g[{\cal P},{\cal Q}] =
\exp
\frac{i}{g}(\sum_{s=0}^{l-1}({\cal P}_s\hat{V}_s
- {\cal Q}^{(s)}\hat{P}_s )).
$$
Substituting vector \r{33} to eq.\r{15}, making use of the relations
$$
\matrix{
(U_g[{\cal P},{\cal Q}])^+ \hat{Q}^{(s)}
U_g[{\cal P},{\cal Q}] =
\hat{Q}^{(s)}+ \frac{1}{g}{\cal Q}^{(s)},
\cr
(U_g[{\cal P},{\cal Q}])^+ \hat{P}_s
U_g[{\cal P},{\cal Q}] =
\hat{P}_s + \frac{1}{g}{\cal P}_s,
}
\l{34}
$$
we obtain that the number
$S_t$ is the action on the classical trajectory satisfying
eqs.\r{25} as $g=0$:
$$
\matrix{
\quad \dot{V}_j = V_{j+1}, j=\overline{0,l-1};
\dot{V}_{l-1} = (-1)^{l-1} \frac{P_{l-1}}{z_l};
\cr
-\dot{P}_j = (-1)^j (z_j+\delta z_j)V_j + P_{j-1}, j=\overline{1,l-1};
\cr
-\dot{P}_0 = (z_0+\delta z_0)Q.
}
\l{y1}
$$
At $g\to 0$ the vector $X_t$ obeys the equation:
$$
i\dot{X}_t =
[H_Q + H_q + \sum_{k=1}^{\infty} \mu_k {\cal Q} q_k + E_0]
X_t.
$$
for a $c$-number $E_0$.
The operator entering to the  right-hand  side  of  this  equation  is
presented as  a  sum  of  the  term corresponding to the field $Q$ and
Hamiltonian \r{a3*}.

Note that the classical solution ${\cal Q}(t)$ can be expressed  as  a
linear combination of the initial conditions for the system \r{y1}:
$$
{\cal Q}(t) = \sum_{s=0}^{l-1} (a_s(t){\cal Q}_s + b_s(t){\cal P}_s)
\l{z4}
$$
Analogously, the Heisenberg operators
$\hat{V}_s(t)=e^{iH_Qt}\hat{V}_se^{-iH_Qt}$ and
$\hat{P}_s(t)=e^{iH_Qt}\hat{P}_se^{-iH_Qt}$ obey the
analog of the system \r{y1} for operator functions. Therefore,
for $\hat{V}_0=\hat{Q}$ we have
$$
\hat{Q}(t) = \sum_{s=0}^{l-1} (a_s(t)\hat{V}_s + b_s(t)\hat{P}_s).
\l{z5}
$$
Eqs. \r{z4} and \r{z5} imply that
$$
U_g^+[{\cal P},{\cal Q}]
g\hat{Q}(t)
U_g[{\cal P},{\cal Q}]
= {\cal Q}(t) + O(g),
\l{z6}
$$
The property \r{z6} will be used in the next subsection.

\subsection{The $\Lambda\to\infty$-limit}

Consider the  limit  $\Lambda\to\infty$.  According to section 2,  the
vector $\Psi_{\Lambda}^t$ should depend on $\Lambda$
according to \r{16}, while $\Phi_{\Lambda}^t$ should be regular as
$\Lambda\to\infty$. Without loss of  generality,  consider  the  fixed
moment of time $t$; index $t$ will be omitted.
Construct the state in the external field obeying the condition
\r{16}:
$$
T_{\Lambda} e^{\frac{i}{g^2} S} U_g[{\cal P},{\cal Q}]Y_{\Lambda}.
\l{i2}
$$
The vector $Y_{\Lambda}$ is regular as
$\Lambda\to \infty$,
$Y_{\Lambda} \to_{\Lambda\to\infty} Y$. Show that for finite values of
$\Lambda$, the  vector  \r{i2}  is  of  the  type \r{33} and therefore
corresponds to the external field.  Comparing eqs.  \r{i2} and \r{33},
one obtains:
$$
X_{\Lambda} = T_{\Lambda}({\cal P},{\cal Q}) Y_{\Lambda},
$$
where
$$
T_{\Lambda}({\cal P},{\cal Q}) = U_g^+[{\cal P},{\cal Q}]
T_{\Lambda} U_g[{\cal P},{\cal Q}].
\l{i3}
$$
The vector  $Y$
can be viewed as a renormalized state in the external  field.  Let  us
check that the operator
$T_{\Lambda}({\cal P},{\cal Q})$ is regular as
$\Lambda=const$, $g\to 0$.

It follows  from eqs.\r{20} and \r{20*} that it transforms the initial
condition for the equation
$$
i \frac{d\tilde{\Phi}^t_{\Lambda}}{dt} =
 U_g^+[{\cal P},{\cal Q}]
 e^{i\hat{H}_0 t}
(g\xi_-(t)\hat{H}_1^{\Lambda}+\hat{H}_{ct}^{\Lambda}[t,g(\cdot)])
e^{-i\hat{H}_0t}
U_g[{\cal P},{\cal Q}]
\tilde{\Phi}^t_{\Lambda}
\l{i4}
$$
at $t=-\infty$ to the solution of this equation as
$t=0$,
$$
\tilde{\Phi}_{\Lambda}^0 =
T_{\Lambda}({\cal P},{\cal Q})
\tilde{\Phi}_{\Lambda}^{-\infty}.
$$
Making use of eq.\r{23}, take eq.\r{i4} to the form
$$
\begin{array}{c}
i \frac{d\tilde{\Phi}^t_{\Lambda}}{dt} =
 U_g^+[{\cal P},{\cal Q}]
(
g\xi_-(t) \sum_k \mu_k \hat{Q}(t)
\frac{\hat{a}_k^+e^{i\Omega_kt}+ \hat{a}_k^- e^{-i\Omega_kt } }
{\sqrt{2\Omega_k}}
\\
+ \sum_{s=0}^{l-1} \frac{(-1)^s}{2} \delta z_s (\hat{Q}^{(s)}(t))^2
)
U_g[{\cal P},{\cal Q}]
\tilde{\Phi}^t_{\Lambda}
\end{array}
\l{i5}
$$
In the leading order in $g$ one has
$$
i \frac{d\tilde{\Phi}^t_{\Lambda}}{dt} =
( \xi_-(t) \sum_k \mu_k {\cal Q}(t)
\frac{\hat{a}_k^+e^{i\Omega_kt}+ \hat{a}_k^- e^{-i\Omega_kt } }
{\sqrt{2\Omega_k}}
+ \sum_{s=0}^{l-1} \frac{(-1)^s}{2g^2} \delta z_s ({\cal Q}^{(s)}(t))^2
)
\tilde{\Phi}^t_{\Lambda}
$$
The evolution operator
$V_{t,-\infty}$
for this equation
which transforms the initial state at
$t=-\infty$  to the solution has the form
$$
V_{t,-\infty} = c_t \exp[\sum_k (\alpha_k(t)\hat{a}_k^+
- \alpha^*_k(t)\hat{a}_k^-)]
\l{i7*}
$$
where
$$
\alpha_k^{\Lambda} (t) = -i \int_{-\infty}^t d\tau  \xi_-(\tau)  {\cal
Q}(\tau) \frac{\mu_k^{\Lambda}}{\sqrt{2\Omega_k}}e^{i\Omega_k \tau}
\l{i7+}
$$
$$
c_t = \exp (\int_{-\infty}^t d\tau
[
-\frac{i}{g^2} \sum_{s=0}^{l-1} \frac{(-1)^s}{2} \delta z_s
({\cal Q}^{(s)}(\tau))^2
+ \frac{1}{2} (\alpha_k^*(\tau)\dot{\alpha}_k(\tau)
- \dot{\alpha}_k^*(\tau){\alpha}_k(\tau) )
] )
$$

In the leading order in $g$
$T_{\Lambda}({\cal P},{\cal Q})= V_{0,-\infty}$.
Note that at
$\Lambda\to \infty$ the operator  \r{i7*}
may be singular, since the series
$\sum_k |\lim_{\Lambda\to\infty} \alpha^{\Lambda}_k|^2$ may diverge.

\subsection{Representations of CCR in the external field}

The average \r{v3} for the semiclassical state \r{i2} has the form:
$$
<e^{\sum_k (z_k \hat{a}_k^+ - z_k^* \hat{a}_k^-)}>=
(Y,
e^{\sum_k (z_k  \pi_{{\cal  P},{\cal  Q}}(a_k^+)  -  z_k^*   \pi_{{\cal
P},{\cal Q}}({a}_k^-))}
Y).
\l{z7}
$$
where
$$
\pi_{{\cal P},{\cal Q}} (a_k^{\pm}) = (T_{\Lambda}(
{\cal P},{\cal Q}))^+ \hat{a}_k^{\pm}
(T_{\Lambda}({\cal P},{\cal Q})).
\l{38}
$$
It follows from eq.\r{i7*} that
$$
\pi_{{\cal P},{\cal     Q}}(a_k^{+})     =     \hat{a}_k^{+}     +
\alpha_k^*,\qquad
\pi_{{\cal P},{\cal Q}}(a_k^{-}) = \hat{a}_k^{-} + \alpha_k
$$
for $\alpha_k$  of  the form \r{i7+} as $\Lambda=\infty$.  We see that
the function \r{z7} is an $\alpha$-density. Integrating eq. \r{i7+} by
parts, we obtain that
$$
[\alpha_k]=
[ - \frac{\mu_k}{\sqrt{2\Omega_k}} \sum_{s=0}^{l-2}
i^s {\cal Q}^{(s)} (0) \frac{1}{\Omega_k^{s+1}}],
$$
provided that the condition \r{13} is satisfied.

This representation coincides with eq.\r{z1}.  We see from the  direct
analysis of  the  exact quantum model that one should choose only such
initial condition for  the  representation  that  relation  \r{z1}  is
satisfied.

It is interesting to note that all the representations
$\pi_{{\cal P},{\cal Q}}$ are {\it equivalent}.
Namely,
$$
\pi_{{\cal P},{\cal Q}} (a_k^{\pm}) =
U_g^+[{\cal P},{\cal Q}]
\pi_{0} (a_k^{\pm})
U_g[{\cal P},{\cal Q}]
\l{i8}
$$
However, the  operator $U_g$ is singular in $g$ and does not possess a
limit $g\to 0$.  This means that representations of CCR appears to  be
equivalent in  the  exact theory and not equivalent in the approximate
theory.

This fact can be understood as follows.  According to subsection  2.3,
the structure of
$\pi_{0} (a_k^{\pm})$ is
$$
\pi_0({a}_k^{-}) =       \hat{a}_k^-       +       \sum_{m=1}^{\infty}
(A_{km}\hat{a}_m^{-} + B_{km}\hat{a}_m^+ )
+ \sum_{s=0}^{l-1} (C_{ks}\hat{V}_s + D_{ks} \hat{P}_s)
$$
for some coefficients
$A_{km}$, $B_{km}$, $C_{ks}$, $D_{ks}$. It follows from section 2 that
at fixed $k$ the vectors with components
$A_{km}$,   $B_{km}$,   $C_{ks}$,  $D_{ks}$ are regular as
$\Lambda\to\infty$.  Consider the limit
$\Lambda\to\infty$.
Eqs.\r{24} imply that the coefficients  $C_{km}$  and  $D_{km}$
are of order  $g$,  while  $A_{kl}$ and $B_{kl}$ are of the order
$g^2$. It follows from eq.\r{34} that the representations
\r{i8} have the following structure:
$$
\pi_{{\cal P},{\cal Q}} (a_k^-) =
\hat{a}_k^- + \beta_k +O(g)
\l{i9}
$$
where
$
\beta_k =
g^{-1}\sum_{s=0}^{l-1}
(C_{ks}{\cal Q}^{(s)} + D_{ks} {\cal P}_s) +O(g)
$
Eqs.\r{i7*}  and  \r{38} imply that $\beta_k$ is
$$
\beta_k=\alpha_k(0).
$$
Eq. \r{i9} is in agreement with obtained in section 3.

Thus, the representations \r{i8} viewed in the leading  order  in  $g$
are nonequivalent if
$\sum_{k}|\beta_k|^2=\infty$
(i.e. if the condition \r{9} is not satisfied).
However, in the exact theory they are equivalent and related with  the
help of the unitary operator which
is singular in $g$.

One can expect that analogous difficulties corresponding to extracting
the classical  component  of  the  field  arise in QED.

This work  was supported by the Russian Foundation for Basic Research,
project 99-01-01198.

\section*{Appendix}

In this appendix the auxiliary mathematical statements are presented.

{\bf Lemma 1}.
{\it
Let $U_n$ be a sequence of unitary operators
in the Fock space, and the sequence of their Wich symbols
$U_n(z^*,w)$ converges to 1 as $n\to\infty$ at arbitrary
$z,w\in  l^2$. Then
$U_n\to 1$ in strong sense.
}

{\sl Proof}.

It follows from the Banach-Shteingaus (?) theorem
(see, for example, \c{22})
that it is sufficient to check that at some
dense subset of the Fock space
$$
||U_nf - f|| \to_{n\to\infty} 0.
\l{a2}
$$
This subset  is  chosen  as a set of all finite linear combinations of
the coherent states
$$
f = \sum_{i=1}^K \alpha_i \exp (\sum_{m=1}^{\infty} z^m_i B^+_m)|0>.
\l{a3}
$$

It follows from the unitarity of the
operator $U_n$ that
for vector \r{a3} the property \r{a2} takes the form
$$
\sum_{ij=1}^K \alpha_i^*    \alpha_j     \exp     (\sum_{m=1}^{\infty}
z^{m*}_iz^m_j) (2-U_n^*(z_i^*,z_j)-U(z_i^*,z_j)) \to_{n\to\infty} 0,
$$
It is satisfied since the Wick symbol converges.

{\bf Corollary 1}.
{\it
Let $U_n$  be  a sequence of unitary operators corresponding to linear
canonical transformations
$$
U_n^{-1} B_m^+ U_n =
\sum_{k=1}^{\infty}( (G_n)_{mk}B_k^+ + (F_n^*)_{mk} B_k^-),
\l{a5}
$$
the operators $F_n$  and  $G_n$ with matrices  $(F_n)_{mk}$  and  $(G_n)_{mk}$
converge as $n\to\infty$ in the operator norm
$
F_n \to_{n\to\infty} 0$ ,
$G_n \to_{n\to\infty} 1$;
for some vectors of the Fock space $f$ and $g$ $(f,U_n g) \to
(f,g)$. Then $U_n \to 1$ in a strong sense.
}

{\sl Proof}.
According to  \c{5}, the Wick symbol of the operator
$U_n$ has the form \r{a1} up to a multiplier.
It follows from lemma 1 that
$\frac{U_n}{<0|U_n|0>} \to 1$. Therefore,
$\frac{(f,U_ng)}{<0|U_n|0>} 
\to \frac{(f,g)}{<0|U_n|0>}$,
 $<0|U_n|0> \to 1$ and $U_n\to 1$.

{\bf Corollary 2}.
{\it
Let $U_n$ be a sequence of unitary operators corresponding  to  linear
canonical transformations \r{a5}, the unitary operator $U$ corresponds
to the canonical transformation
$$
\begin{array}{c}
U^{-1} B_m^+ U =
\sum_{k=1}^{\infty}( G_{mk}B_k^+ + F^*_{mk} B_k^-),
\\
F_n \to_{n\to\infty} F ,
G_n \to_{n\to\infty} G; <0|U_n|0> \to <0|U|0>.
\end{array}
$$
Then $U_n \to 1 $ in a strong sense.
}

{\bf Corollary 3}.
{\it
Let $U_n$ be a sequence of unitary operators
$$
U_n = \exp(\sum_{k=1}^{\infty} (z_k^{(n)}
\hat{a}_k^+ - z^{(n)*}_k \hat{a}_k^-)),
$$
while
$$
U = \exp(\sum_{k=1}^{\infty} (z_k
\hat{a}_k^+ - z_k \hat{a}_k^-)),
$$
where $z^{(n)} \in l^2$, $z \in l^2$, $||z^{(n)}-z||\to_{n\to\infty} 0$.
Then $U_n \to 1 $ in a strong sense.
}

To prove  the corollaries,  it is sufficient to consider the sequence of
unitary operators
$U_nU^{-1}$ and use the statement of corollary 1.

{\bf Lemma 2}.
{\it
Let $A_n$ and $B_n$ be Hilbert-Schmidt operators, which converge in the
norm \r{k8} to operators $A$ and $B$, while the operator
$1+AB$ is invertible. Then
$
det(1+A_nB_n) \to_{n\to\infty} det (1+AB)
$.
}

{\sl Proof}.
The property \r{22} $Tr|XYZ| \le ||X||_2||Y|| ||Z||_2$ imply that  for
the quantity
$
c_n =
\frac{det(1+A_nB_n)}{ det (1+AB)}
$
$$
\begin{array}{c}
|ln c_n| = |\int_0^1 d\tau Tr
([1+\tau(1+AB)^{-1}(A_nB_n-AB)]^{-1}
\\
(1+AB)^{-1}(A_nB_n-AB)) |
\le
\\
max_{\tau \in (0,1)}
||[1+\tau(1+AB)^{-1}(A_nB_n-AB)]^{-1}||
||(1+AB)^{-1}||
\\
(||A_n-A||_2 ||B_n||_2+ ||A||_2||B_n-B||_2) \to_{n\to\infty} 0.
\end{array}
$$
Lemma is proved.

Consider the sequence of the Volterra integral equations
$$
q_n(t) = \alpha_n(t) + \int_{-T}^t d\tau f_n(t,\tau) q_n(\tau),
$$
for $q_n$  at  $[-T,T]$.
Denote by
$G_n(t,\tau)$ the Green functions for these equations
which are defined form the relation
$$
q_n(t) = \int_{-T}^t G_n(t,\tau) \alpha_n(\tau) d\tau
$$

{\bf Lemma  3}.
{\it
Let the sequences of functions
$f_n$ and $\alpha_n$ uniformly converge to $f$ and $\alpha$. Then
the sequence $G_n(t,\tau)$
uniformly converge to the Green function of the equation
$$
q(t) = \alpha(t) + \int_{-T}^t d\tau f(t,\tau) q(\tau),
$$
and the sequence $q_n$ uniformly converge to $q$.
}

To prove  the lemma,  it is sufficient to use the explicit form of the
Green function which is obtained from the iteration procedure of
\c{18}.

\newpage
\pagestyle{empty}

\newpage

\begin{picture}(100,100)
\qbezier(10,10)(20,10)(30,30)
\qbezier(50,50)(40,50)(30,30)
\put(50,50){\line(1,0){10}}
\put(10,10){\line(-1,0){10}}
\put(0,10){\line(1,0){100}}
\put(100,10){\line(-2,1){4}}
\put(100,10){\line(-2,-1){4}}
\put(10,10){\line(0,1){5}}
\put(50,10){\line(0,1){5}}
\put(50,20){\line(0,1){5}}
\put(50,30){\line(0,1){5}}
\put(50,45){\line(0,1){5}}
\put(60,0){\line(0,1){100}}
\put(60,100){\line(1,-2){2}}
\put(60,100){\line(-1,-2){2}}
\put(50,50){\line(-1,0){5}}
\put(40,50){\line(-1,0){5}}
\put(30,50){\line(-1,0){5}}
\put(20,50){\line(-1,0){5}}
\put(10,50){\line(-1,0){5}}
\put(63,95){$\xi_0$}
\put(61,50){$1$}
\put(61,12){$0$}
\put(95,5){$\tau$}
\put(5,0){$-T_1$}
\put(45,0){$-T_2$}
\put(50,-20){Fig.1}
\end{picture}

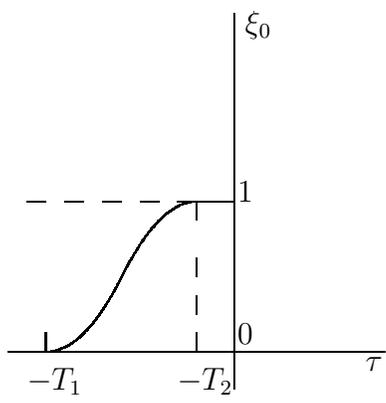
\end{document}